\documentclass[numsec,webpdf,modern,medium,namedate]{oup-authoring-template}

\onecolumn

\graphicspath{{Fig/}}

\theoremstyle{thmstyleone}%
\theoremstyle{thmstyletwo}%
\theoremstyle{thmstylethree}%

\usepackage[doublespacing]{setspace}
\usepackage[fontsize=12pt]{fontsize}

\begin{document}

\journaltitle{Journals of the Royal Statistical Society}
\DOI{DOI HERE}
\copyrightyear{XXXX}
\pubyear{XXXX}
\access{Advance Access Publication Date: Day Month Year}
\appnotes{Original article}

\firstpage{1}


\title[TEASE]{TEASE: Targeted elastic spatial envelope}

\author[1,$\ast$]{Mitchell L.\ Krock}

\authormark{Krock}

\address[1]{\orgdiv{Department of Statistics and Data Science}, \orgname{University of Missouri}, \orgaddress{\street{146 Middlebush Hall}, \postcode{65211}, \state{Missouri}, \country{USA}}}

\corresp[$\ast$]{Department of Statistics and Data Science, University of Missouri, 146 Middlebush Hall, Columbia, MO, 65211, USA \href{Email:mk52n@missouri.edu}{mk52n@missouri.edu}}

\received{Date}{0}{Year}
\revised{Date}{0}{Year}
\accepted{Date}{0}{Year}



\abstract{
Envelope regression is a crucial part of multivariate linear modeling, leveraging separation into material and immaterial parts to provide parsimonious data reduction.
\citet{rekabdarkolaee2020} construct spatial envelope, a novel multivariate Gaussian process, extended by \citet{may2022} to linear coregionalization envelope.
Our proposal (TEASE) combines spatial envelope and full-scale basis graphical lasso \citep{leduc2025modelinglargenonstationaryspatial}, leveraging 
targeted graphical elastic net \citep{kovács2021graphicalelasticnettarget} for material precision matrices.
Immaterial part is a multivariate multiresolution Gaussian Markov random field \citep{kleiber2019,caringi2026fixedrankcokrigingmodel}.
Enhanced envelope penalty \citep{kwon2024} regularizes regression coefficients.
Demonstrations are on climate data.}


\maketitle

\section{Introduction}
\label{sec:introduction}

\cite{cook2010} seminally introduced envelopes for multivariate linear models.
By projecting response variables into material and immaterial parts, envelopes are useful for dimension reduction \citep{soale2026}, and can provide drastic efficiency gains over classical regression, especially if immaterial envelope explains significantly more variance than material.
Variants include partial envelope \citep{su2011, ZhangJing2023SPEM}, scaled envelope \citep{cook2013,zhang2024}, quantile envelope \citep{ding2021, lee2022, zhao2022}, Bayesian envelope \citep{khare2017,Chakraborty02072024,zeng2025bayesianapproachrobustlongitudinal,kim2026laplacevariationalinferencebayesian}, 
tensor envelope \citep{Li03072017,naji2026bayesiantensorregressionneuroimaging},
sparse envelope \citep{su2016}, and
nonparametric envelope \citep{jacobson2025nonparametricenvelopesflexibleresponse}.
Envelopes can handle ignorable missing data \citep{ma2021}, and are connected to principal components \citep{jacobson2025envelopeguidedregularizationimprovedprediction,zhang2023} and partial least squares \citep{cook2020, wu2025envelopebasedpartialsquaresfunctional,cook2013partial}.
\cite{franks2022} allows mean and covariance to depend upon covariates.
\cite{han2018robustnessshaperestrictedregressionestimators} leverage envelopes to characterize robustness of shape-restricted regression.
\cite{ding2018} introduce matrix-variate envelope regression.
See \cite{cook2015} for envelope foundations and \cite{lee2020} to review models.

\cite{rekabdarkolaee2020} created spatial envelope for multivariate Gaussian Processes (GPs).
Also called multi-task or multi-output, multivariate GPs are popular in statistics and machine learning \citep{bruinsma2020,zhang2025physicsaugmentedmultitaskgaussianprocess,gninkou2026scalablemultitaskgaussianprocesses,gevorgyan2026exchangeablegaussianprocessesstaggeredadoption}; \cite{genton2015} review classical methods, including linear coregionalization model (LCM).
After introducing envelope regression (Section \ref{sec:envelope}), we discuss spatial and LCM envelopes \citep{may2022} in Section \ref{sec:spatialenvelope}.
Targeted elastic spatial envelope, described in Sections \ref{sec:bgl} and \ref{sec:opt}, is deployed on climate data in Section \ref{sec:data}. 
Section \ref{sec:discussion} concludes.

\section{Envelope}
\label{sec:envelope}

  \cite{cook2010} consider the linear model
  \begin{equation}
  \label{eq:linearmodel}
  \mathbf{Y} =\boldsymbol{\alpha} + \boldsymbol{\beta} \mathbf{X} + \boldsymbol{\epsilon}.
  \end{equation}  
$\mathbf{Y} \in \mathbb{R}^r$ is the response vector, $\boldsymbol{\alpha} \in \mathbb{R}^r$ is a vector of intercepts, $\mathbf{X} \in \mathbb{R}^p$ is a predictor vector, $\boldsymbol{\beta} \in \mathbb{R}^{r, p}$ is a matrix of regression coefficients, and
$\boldsymbol{\epsilon} \sim N(\mathbf{0},\boldsymbol{\Sigma})$ with $\boldsymbol{\Sigma} \in \mathbb{R}^{r,r}_{\succeq 0}$.
Suppose there exists an orthonormal matrix $\boldsymbol{\Gamma} = [\boldsymbol{\Gamma}_1 \ \boldsymbol{\Gamma}_0] \in \mathbb{R}^{r, r}$ with
\begin{enumerate}
  \item $\boldsymbol{\beta} \in \text{span}(\boldsymbol{\Gamma}_1)$,
  \item $\boldsymbol{\Gamma}_0^{\mathrm{T}} \mathbf{Y} \perp \boldsymbol{\Gamma}_1^{\mathrm{T}} \mathbf{Y} \mid \mathbf{X} \iff \text{Cov}(\boldsymbol{\Gamma}_0^{\mathrm{T}} \mathbf{Y}, \boldsymbol{\Gamma}_1^{\mathrm{T}} \mathbf{Y} \mid \mathbf{X}) = \mathbf{0}$.
\end{enumerate}
Then, $\boldsymbol{\Gamma}_0^{\mathrm{T}} \mathbf{Y}$, called immaterial, is marginally independent of $\mathbf{X}$, and conditionally independent of $\mathbf{X}$ given material part $\boldsymbol{\Gamma}_1^{\mathrm{T}} \mathbf{Y}$.
Envelopes reduce the number of regression parameters by $p(r-u)$, where $\boldsymbol{\Gamma}_1$ has $u$ columns.
Note $\boldsymbol{\Sigma} = \mathcal{P}_{\boldsymbol{\Gamma}_0} \boldsymbol{\Sigma} \mathcal{P}_{\boldsymbol{\Gamma}_0} + \mathcal{P}_{\boldsymbol{\Gamma}_1} \boldsymbol{\Sigma} \mathcal{P}_{\boldsymbol{\Gamma}_1}$ splits into immaterial and material parts, where $\mathcal{P}_{\boldsymbol{\Gamma}_1} = \boldsymbol{\Gamma}_1 (\boldsymbol{\Gamma}_1 ^{\mathrm{T}} \boldsymbol{\Gamma}_1)^{-1} \boldsymbol{\Gamma}_1^{\mathrm{T}}$ is a projection operator. 
Estimation of $\boldsymbol{\Gamma}_1$ and $\boldsymbol{\Gamma}_0$ involves Riemannian optimization on the Stiefel manifold \citep{edelman1998}.
Researchers are keen on developing fast algorithms for envelopes, often obviating Stiefel orthogonality constraints \citep{cook2016algorithms,cook2016fast,zhang2023}. 
\cite{cookbook} is a textbook overview of envelope methods.

\section{Spatial Envelope}
\label{sec:spatialenvelope}

\cite{rekabdarkolaee2020} propose spatial envelope for $r$-variate spatial process $\mathbf{Y}(\mathbf{s}) = (Y_1(\mathbf{s}),\dots,Y_r(\mathbf{s}))^{\mathrm{T}}$ and $p$ predictors $\mathbf{X}(\mathbf{s}) = (X_1(\mathbf{s}),\dots,X_p(\mathbf{s}))^{\mathrm{T}}$ at locations $\mathbf{s}_1,\dots,\mathbf{s}_n$. Process model in matrix form is
\begin{equation}
    \label{eq:rekabdarkolaee2020mat}
  [\mathbf{Y}(\mathbf{s}_1) \mathbf{Y}(\mathbf{s}_2) \cdots \mathbf{Y}(\mathbf{s}_n)] = \boldsymbol{\alpha} \otimes \mathbf{1}_n^{\mathrm{T}} + \boldsymbol{\Gamma}_1 \boldsymbol{\eta} [\mathbf{X}(\mathbf{s}_1) \mathbf{X}(\mathbf{s}_2) \cdots \mathbf{X}(\mathbf{s}_n)] + \boldsymbol{\epsilon}(\mathbf{s}),
\end{equation}
where $\boldsymbol \eta \in \mathbb{R}^{u,p}$ and $\boldsymbol{\epsilon}({\mathbf{s}})$ is a mean zero GP such that
\begin{equation}
  \label{eq:vsplitmaterial}
\text{Cov}(\boldsymbol{\epsilon}(\mathbf{s}_i), \boldsymbol{\epsilon}(\mathbf{s}_j)) = \mathbf{V} \rho(\| \mathbf{s}_i - \mathbf{s}_j\|), \quad \mathbf{V} = \boldsymbol{\Gamma}_1 \boldsymbol{\Omega}_1 \boldsymbol{\Gamma}_1^{\mathrm{T}} + \boldsymbol{\Gamma}_0 \boldsymbol \Omega_0 \boldsymbol{\Gamma}_0^{\mathrm{T}},
\end{equation}
with $\rho(\cdot)$ an isotropic correlation function (e.g.,\ Matérn) shared among the $r$ responses.
That is, the covariance matrix is partitioned into material and immaterial parts with respective bases $\boldsymbol{\Gamma}_1 \in \mathbb{R}^{r,u}$ and $\boldsymbol{\Gamma}_0 \in \mathbb{R}^{r,r-u}$, and the regression coefficient matrix lies in the column space of $\boldsymbol{\Gamma}_1$.
Spatial envelope shows substantial improvement over multivariate linear regression, LCM, envelope, and spatial envelope in an analysis of 8 variables (ground level ozone, sulfur dioxide, carbon monoxide, nitrogen dioxide, nitrogen monoxide, lead, PM 2.5, PM 10) using 5 predictors (wind, temperature, relative humidity, latitude, longitude) at 270 air monitors in NE USA.

\subsection{Linear coregionalization envelope}
\label{subsec:lmc}

LCM has a rich history in multivariate spatial statistics \citep{goulard1992,schmidt2003,gelfand2004,zhang_spatialfactor,liu2022,guinnessmultivar,guhaniyogi2013,jeon2026uncertaintyawareneuralmultivariategeostatistics,ROY2026100948,stevenson2026gaussianprocesslatentfactor}.
Linear computational complexity in response dimension permits rapid computations with large datasets \citep{alie2024computationalconsiderationslinearmodel,truffinet2024exactgeneraldecoupledsolutions}.
\cite{may2022} introduce LCM envelope, rewriting the right hand side of \eqref{eq:rekabdarkolaee2020mat} as
\begin{equation}
    \label{eq:may2022mat}
\boldsymbol{\alpha} \otimes \mathbf{1}_n^{\mathrm{T}} + \boldsymbol{\Gamma}_1 \boldsymbol{\eta} [\mathbf{X}(\mathbf{s}_1) \mathbf{X}(\mathbf{s}_2) \cdots \mathbf{X}(\mathbf{s}_n)] + \boldsymbol{\Gamma}_1 \mathbf{D}_1^{1/2} \boldsymbol{\epsilon}_1(\mathbf{s}) + \boldsymbol{\Gamma}_0 \mathbf{D}_0^{1/2}\boldsymbol{\epsilon}_0(\mathbf{s})
\end{equation}
where $\mathbf{D}_1, \mathbf{D}_0$ are diagonal with elements $\omega_1,\dots,\omega_r$
so that $\text{Cov}(\mathbf{Y}(\mathbf{s}_i), \mathbf{Y}(\mathbf{s}_j)) = \boldsymbol{\Gamma}_1 \text{diag}(\omega_1 \rho_1(d_{ij}), \dots, \omega_u \rho_u(d_{ij}))  \boldsymbol{\Gamma}_1^{\mathrm{T}}
+ \boldsymbol{\Gamma}_0 \text{diag}(\omega_{u+1} \rho_{u+1}(d_{ij}), \dots, \omega_r \rho_r(d_{ij}))  \boldsymbol{\Gamma}_0^{\mathrm{T}}$, $d_{ij} = \|\mathbf{s}_i - \mathbf{s}_j\|$.
By design, LCM envelope has a non-separable covariance.
\citet{may2022} study covariate (nitrogen, carbon) effects on responses (phosphorous, potassium, zinc, soil PH level, nitrate, ammonium) and find LCM envelope has lower BIC than classical LCM and spatial envelope.

\section{TEASE (Targeted ElAstic Spatial Envelope)}
\label{sec:bgl}

Basis models are a fundamental part of geostatistics \citep{cressie2022}.
\cite{ekanayaka2025} leverage basis graphical lasso (BGL) \citep{krock2023mbgl,krock2021bgl} for state-of-the-art downscaled Great Barrier Reef sea surface temperature projections of CMIP6 Earth system models.
Full-scale BGL \citep{leduc2025modelinglargenonstationaryspatial} leverages full-scale approximation \citep{sang2012,stein2008,zhang2015,zhang2019,gyger2026,beckman2023}, improving performance in small scale variability.
\cite{chen2026penalizedlikelihoodinferencehighly} propose an idea similar to BGL but in spectral domain.

For targeted elastic spatial envelope, the right hand side of \eqref{eq:rekabdarkolaee2020mat} reads
\begin{equation}
  \label{eq:processmodel}
  \boldsymbol{\alpha} \otimes \mathbf{1}_n^{\mathrm{T}} + \boldsymbol{\Gamma}_1 \boldsymbol{\eta} [\mathbf{X}(\mathbf{s}_1) \mathbf{X}(\mathbf{s}_2) \cdots \mathbf{X}(\mathbf{s}_n)] + \boldsymbol{\Gamma}_1 [\mathbf{W}_1 \cdots \mathbf{W}_\ell] \boldsymbol \Phi_1^{\mathrm{T}} + \boldsymbol{\Gamma}_0 [\mathbf{V}_1 \cdots \mathbf{V}_m] \boldsymbol \Phi_0^{\mathrm{T}}.
\end{equation}
Whereas LCM is a linear combination of independent Gaussian processes, \eqref{eq:processmodel} is a linear combination of Gaussian random vectors.
Basis matrices $\boldsymbol{\Phi}_0 \in \mathbb{R}^{n,m}$ and $\boldsymbol{\Phi}_1 \in \mathbb{R}^{n,\ell}$ are row-indexed by spatial location and column-indexed by basis function, i.e.,\ entry $(i,j)$ is $\varphi_j(\mathbf{s}_i)$, the $j$th basis function evaluated at $\mathbf{s}_i$.

Let $\mathbf{W} = (\mathbf{W}_1^{\mathrm{T}},\dots,\mathbf{W}_\ell^{\mathrm{T}})^{\mathrm{T}}$ and $\mathbf{V} = (\mathbf{V}_1^{\mathrm{T}},\dots,\mathbf{V}_m^{\mathrm{T}})^{\mathrm{T}}$ be independent with mean zero and respective precision matrices $\mathbf{Q}_W, \mathbf{Q}_V.$
Columns of $\mathbf{W}$ are independent Gaussian graphical vectors. 
Write block diagonal matrix $\mathbf{Q}_W = \bigoplus_{i=1}^\ell \mathbf{Q}_{W,i}$, $\text{Cov}(\mathbf{W}_i) = \mathbf{Q}_{W,i}^{-1} \in \mathbb{R}_{\succeq 0}^{u,u}$, corresponding to $\boldsymbol{\Phi}_1$, a low rank matrix of EOFs.
Columns of $\mathbf{V}$ form a multiresolution Gaussian Markov random field (GMRF) with $m_\ell$ levels of resolution; see \citet{rue2005} for GMRF theory and applications.
Inspired by fixed rank co-kriging \citep{caringi2026fixedrankcokrigingmodel}, $\boldsymbol{\Phi}_0$ is a sparse matrix of compactly-supported multiresolution Wendland basis functions associated with block diagonal precision matrix $
\mathbf{Q}_V= \bigoplus_{h=1}^{m_\ell} \mathbf{Q}_{V,h} \otimes (\mathbf{B}_h \mathbf{B}_h^{\mathrm{T}})$.
Multiresolution Wendland bases matrix $\boldsymbol{\Phi}_0$ and SAR matrices $\{\mathbf{B}_h\}_{h=1}^{m_\ell}$ are constructed from LatticeKrig \citep{nychka2015}.
$\mathbf{B}_h$ is controlled by one parameter $\kappa_0^2$ centering a finite difference approximation with stencil 
\begin{equation} \label{e:2}
  \begin{array}{c|c|c}
    0 & -1 & 0 \\
    \hline
    -1 & 4+ \kappa_0^2 & -1 \\
    \hline
    0 & -1 & 0 \\
  \end{array}
\end{equation}
LatticeVision \citep{sikorski2026latticevisionimageimagenetworks,moro2026nonstationaryamortizedtransferlearning} leverages image-to-image neural networks for SAR inference in nonstationary LatticeKrig.

\subsection{Likelihood}
\label{subsec:likelihood}

Vectorizing \eqref{eq:processmodel} with identity $\displaystyle \left(\mathbf {B} ^{\mathrm {T}}\otimes \mathbf {A} \right)\,\operatorname {vec} (\mathbf {X} )=\operatorname {vec} (\mathbf {AXB} )=\operatorname {vec} (\mathbf {C} )$:
\begin{equation}
\label{eq:vecmodel}  
\begin{pmatrix}
\mathbf{Y}(\mathbf{s}_1) \\
\mathbf{Y}(\mathbf{s}_2) \\
\vdots \\
\mathbf{Y}(\mathbf{s}_n)
\end{pmatrix}  = \mathbf{1}_n \otimes \boldsymbol{\alpha}  + (\mathbf{I}_n \otimes \boldsymbol{\Gamma}_1 \boldsymbol{\eta}) \begin{pmatrix}
  \mathbf{X}(\mathbf{s}_1) \\
  \mathbf{X}(\mathbf{s}_2) \\
  \vdots \\
  \mathbf{X}(\mathbf{s}_n)
  \end{pmatrix} 
  + (\boldsymbol{\Phi}_1 \otimes \boldsymbol{\Gamma}_1) 
  \begin{pmatrix}
    \mathbf{W}_1 \\
    \mathbf{W}_2 \\
    \vdots \\
    \mathbf{W}_\ell
    \end{pmatrix}
      + (\boldsymbol{\Phi}_0 \otimes \boldsymbol{\Gamma}_0) 
  \begin{pmatrix}
    \mathbf{V}_1 \\
    \mathbf{V}_2 \\
    \vdots \\
    \mathbf{V}_m
    \end{pmatrix}.
  \end{equation}

Suppose $\mathbb{Y}_1,\dots,\mathbb{Y}_t$ are realizations of $(\mathbf{Y}(\mathbf{s}_1)^{\mathrm{T}},\dots,\mathbf{Y}(\mathbf{s}_n)^{\mathrm{T}})^{\mathrm{T}}$.
Also let $\mathbb{X}_k = (\mathbf{X}(\mathbf{s}_1)^{\mathrm{T}},\dots,\mathbf{X}(\mathbf{s}_n)^{\mathrm{T}})^{\mathrm{T}}$ for $k \in \{ 1,\dots,t \}$, noting the right hand side could change with $k$.
Unpenalized negative log likelihood is $nr \log 2\pi$ plus
\begin{align}
  \label{eq:negloglik1}
  & \log \det \nolimits_0 \left((\boldsymbol{\Phi}_1 \otimes \boldsymbol{\Gamma}_1 ) \mathbf{Q}_W^{-1} (\boldsymbol{\Phi}_1^{\mathrm{T}} \otimes \boldsymbol{\Gamma}_1^{\mathrm{T}} )  + (\boldsymbol{\Phi}_0 \otimes \boldsymbol{\Gamma}_0) \mathbf{Q}_V^{-1} (\boldsymbol{\Phi}_0^{\mathrm{T}} \otimes \boldsymbol{\Gamma}_0^{\mathrm{T}})   \right) \\ \nonumber
  + & \text{tr}\left( \mathbf{S} \left((\boldsymbol{\Phi}_1 \otimes \boldsymbol{\Gamma}_1 ) \mathbf{Q}_W^{-1} (\boldsymbol{\Phi}_1^{\mathrm{T}} \otimes \boldsymbol{\Gamma}_1^{\mathrm{T}} )  + (\boldsymbol{\Phi}_0 \otimes \boldsymbol{\Gamma}_0) \mathbf{Q}_V^{-1} (\boldsymbol{\Phi}_0^{\mathrm{T}} \otimes \boldsymbol{\Gamma}_0^{\mathrm{T}})   \right)^{-1} \right);
\end{align}
$\mathbf{S} = \dfrac{1}{t} \displaystyle \sum_{k=1}^t (\mathbb{Y}_k-\mathbf{1}_n \otimes \boldsymbol \alpha - (\mathbf{I}_n \otimes \boldsymbol{\Gamma}_1 \boldsymbol{\eta}) \mathbb{X}_k)(\mathbb{Y}_k-\mathbf{1}_n \otimes \boldsymbol \alpha - (\mathbf{I}_n \otimes \boldsymbol{\Gamma}_1 \boldsymbol{\eta}) \mathbb{X}_k)^{\mathrm{T}}$ denotes sample covariance.
Here we introduce the pseudoinverse, noting $\boldsymbol{\Phi}_1^\dagger = \boldsymbol{\Phi}_1^{\mathrm{T}}$ by semi-orthogonality.
Envelope orthogonality motivates use of Lemma 4.2 in \cite{cook2010}: $\det_0(A+B) = \det_0(A) \det_0(B)$ and $(A+B)^\dagger = A^\dagger + B^\dagger$, for nonzero symmetric $A,B$ such that $AB = 0$.
Pseudoinverse of $(\boldsymbol{\Phi}_1 \otimes \boldsymbol{\Gamma}_1 ) \mathbf{Q}_W^{-1} (\boldsymbol{\Phi}_1^{\mathrm{T}} \otimes \boldsymbol{\Gamma}_1^{\mathrm{T}} )  + (\boldsymbol{\Phi}_0 \otimes \boldsymbol{\Gamma}_0 ) \mathbf{Q}_V^{-1} (\boldsymbol{\Phi}_0^{\mathrm{T}} \otimes \boldsymbol{\Gamma}_0^{\mathrm{T}})$ 
equals $(\boldsymbol{\Phi}_1 \otimes \boldsymbol{\Gamma}_1 ) \mathbf{Q}_W (\boldsymbol{\Phi}_1^{\mathrm{T}} \otimes \boldsymbol{\Gamma}_1^{\mathrm{T}} )  + (\boldsymbol{\Psi}_0 \otimes \boldsymbol{\Gamma}_0 ) \mathbf{Q}_V (\boldsymbol{\Psi}_0^{\mathrm{T}} \otimes \boldsymbol{\Gamma}_0^{\mathrm{T}})$, where $\boldsymbol{\Psi}_0^{\mathrm{T}} = \boldsymbol{\Phi}_0^\dagger$.
Thus immaterial and material parts of \eqref{eq:negloglik1} separate:
\begin{align}
  \label{eq:negloglik2}
  & \log \det \nolimits_0 \left((\boldsymbol{\Phi}_1 \otimes \boldsymbol{\Gamma}_1 ) \mathbf{Q}_W^{-1} (\boldsymbol{\Phi}_1^{\mathrm{T}} \otimes \boldsymbol{\Gamma}_1^{\mathrm{T}} ) \right)  +  \log \det \nolimits_0 \left( (\boldsymbol{\Phi}_0 \otimes \boldsymbol{\Gamma}_0) \mathbf{Q}_V^{-1} (\boldsymbol{\Phi}_0^{\mathrm{T}} \otimes \boldsymbol{\Gamma}_0^{\mathrm{T}})   \right) \\ \nonumber
  + & \text{tr}\left( \mathbf{S} \left((\boldsymbol{\Phi}_1 \otimes \boldsymbol{\Gamma}_1 ) \mathbf{Q}_W (\boldsymbol{\Phi}_1^{\mathrm{T}} \otimes \boldsymbol{\Gamma}_1^{\mathrm{T}} ) \right) \right) + \text{tr} \left( \mathbf{S} \left(  (\boldsymbol{\Psi}_0 \otimes \boldsymbol{\Gamma}_0) \mathbf{Q}_V (\boldsymbol{\Psi}_0^{\mathrm{T}} \otimes \boldsymbol{\Gamma}_0^{\mathrm{T}})   \right) \right).
\end{align}

Since $\boldsymbol{\Phi}_1$ and $\boldsymbol{\Gamma}_1$ are semi-orthonormal, material log determinant is $\log \det \mathbf{Q}_W^{-1} = - \sum_{i=1}^\ell \log \det \mathbf{Q}_{W,i}$, while immaterial log determinant equals
\begin{equation}
  \label{eq:immateriallogdet}
\log \det \nolimits_0 \left( ( \boldsymbol{\Phi}_0 \otimes \mathbf{I}_{r-u} )  \left(  \bigoplus_{h=1}^{m_\ell} \mathbf{Q}_{V,h} \otimes (\mathbf{B}_h \mathbf{B}_h^{\mathrm{T}}) \right)^{-1} (\boldsymbol{\Phi}_0^{\mathrm{T}} \otimes \mathbf{I}_{r-u}) \right).
\end{equation}
Let $\tilde{\mathbb{Y}}_k = \mathbb{Y}_k - \mathbf{1}_n \otimes \boldsymbol{\alpha}$ and $\tilde{\mathbf{Y}}_k= \text{mat}(\tilde{\mathbb{Y}}_k)$ and $\mathbf{X}_k = \text{mat}(\mathbb{X}_k)$.
Material trace equals
$\text{tr} \left( \mathbf{S}_{1}  \mathbf{Q}_{W} \right) = \sum_{i=1}^\ell \text{tr}(\mathbf{S}_{1,i} \mathbf{Q}_{W,i} ), \\ 
\mathbf{S}_{1} = \dfrac{1}{t} \displaystyle \sum_{k=1}^t \text{vec}\left(\boldsymbol{\Gamma}_1^{\mathrm{T}} \tilde{\mathbf{Y}}_k \boldsymbol{\Phi}_{1}  - \boldsymbol{\eta} {\mathbf{X}}_k \boldsymbol{\Phi}_{1} \right) \text{vec} \left(\boldsymbol{\Gamma}_1^{\mathrm{T}} \tilde{\mathbf{Y}}_k \boldsymbol{\Phi}_{1} - \boldsymbol{\eta} {\mathbf{X}}_k \boldsymbol{\Phi}_{1} \right)^{\mathrm{T}} \in \mathbb{R}^{\ell u,\ell u}_{\succeq 0}$.
Pull out block diagonals to form $\{\mathbf{S}_{1,i}\}_{i=1}^\ell$.
Due to envelope orthogonality, immaterial trace is 
\begin{equation}
\text{tr}(\mathbf{S}_0 \mathbf{Q}_V) = \frac{1}{t} \sum_{k=1}^t (\text{vec}(\boldsymbol{\Gamma}_0^{\mathrm{T}} \tilde{\mathbf{Y}}_{k} \boldsymbol{\Psi}_0))^{\mathrm{T}} \mathbf{Q}_V \text{vec}(\boldsymbol{\Gamma}_0^{\mathrm{T}} \tilde{\mathbf{Y}}_{k} \boldsymbol{\Psi}_0).
\end{equation}
Instead of computing $\boldsymbol{\Psi}_0$, normal equations of sparse linear systems are solved, adding small ridge regression penalty if necessary.

An advantage of TEASE is non-separable covariance:
\begin{align}
&\text{Cov}(\mathbf{Y}(\mathbf{s}), \mathbf{Y}(\mathbf{s}'))
= \sum_{i=1}^\ell \phi_{1,i}(\mathbf{s}) \phi_{1,i}(\mathbf{s}') \boldsymbol{\Gamma}_1 \mathbf{Q}_{W,i}^{-1} \boldsymbol{\Gamma}_1^{\mathrm{T}} \\ \nonumber
&+ \sum_{h=1}^{m_\ell} \left( \left( \boldsymbol{\phi}_{0,h}(\mathbf{s} )\right)^{\mathrm{T}} \otimes \boldsymbol{\Gamma}_0 \right) (\mathbf{Q}_{V,h}^{-1} \otimes (\mathbf{B}_h \mathbf{B}_{h}^{\mathrm{T}})^{-1}) \left(\boldsymbol{\phi}_{0,h}(\mathbf{s}') \otimes \boldsymbol{\Gamma}_0^{\mathrm{T}}\right)
\end{align}
with $\boldsymbol{\phi}_{0,h}(\mathbf{s}) \in \mathbb{R}^{n_h}$ so that, in general, 
$\text{Cov}(\mathbf{Y}_j(\mathbf{s}), \mathbf{Y}_k(\mathbf{s}')) \neq 
\text{Cov}(\mathbf{Y}_k(\mathbf{s}), \mathbf{Y}_j(\mathbf{s}')).
$

\subsection{Penalized Likelihood}
\cite{kwon2024} shrink regression coefficients after the material part is standardized to have uncorrelated errors.
Under linear model \eqref{eq:linearmodel}, $\boldsymbol{\Omega}^{-1/2}_1 \boldsymbol{\Gamma}_1^{\mathrm{T}} \mathbf{Y} = \boldsymbol{\Omega}_1^{-1/2} \boldsymbol{\eta X} + \boldsymbol{\Omega}_1^{-1/2}\boldsymbol{\Gamma}_1^{\mathrm{T}} \boldsymbol{\epsilon}$ means $\boldsymbol{\Omega}^{-1/2}_1  \boldsymbol{\Gamma}_1^{\mathrm{T}} \boldsymbol{\epsilon} \sim N(\mathbf{0},\mathbf{I}_r)$ since $\boldsymbol{\epsilon} \sim N(\mathbf{0}, \boldsymbol{\Gamma}_1 \boldsymbol{\Omega}_1 \boldsymbol{\Gamma}_1^{\mathrm{T}} + \boldsymbol{\Gamma}_0 \boldsymbol{\Omega}_0 \boldsymbol{\Gamma}_0^{\mathrm{T}} )$, motivating penalty $\rho(\boldsymbol \eta, \boldsymbol{\Omega}_1) = \text{tr}(\boldsymbol \eta^{\mathrm{T}} \boldsymbol{\Omega}_1^{-1} \boldsymbol \eta)$.
Enhanced envelope prediction risk is superior to standard envelope.
Moreover, enhanced envelope regularization solves identifiability issues due to Stiefel manifold's invariance under orthogonal transformations. 
TEASE enhanced envelope penalty, weighted with $\mu>0$, equals
\begin{equation}
  \label{eq:envelopepenalty}
  \text{tr}\left((\boldsymbol{\Phi}_1 \otimes  \boldsymbol{\eta}^\mathrm{T} ) \mathbf{Q}_W (\boldsymbol{\Phi}_1^\mathrm{T} \otimes \boldsymbol{\eta}) \right) 
  = \sum_{i=1}^\ell  \text{tr}(\boldsymbol{\eta}^\mathrm{T}  \mathbf{Q}_{W,i} \boldsymbol{\eta}).
\end{equation}
Material precision matrices $\{\mathbf{Q}_{W,i}\}_{i=1}^\ell$ are also regularized with targeted graphical elastic net \citep{kovács2021graphicalelasticnettarget}. 

Collecting terms, TEASE objective equals
\begin{align}  
  \label{eq:finalobjective}
&  \log \det \nolimits_0 \left( ( \boldsymbol{\Phi}_0 \otimes \mathbf{I}_{r-u} ) \mathbf{Q}_{V}^{-1} (\boldsymbol{\Phi}_0^{\mathrm{T}} \otimes \mathbf{I}_{r-u}) \right) +  \text{tr}\left( \mathbf{S}_0 \mathbf{Q}_V \right) +  \\
 & \hspace{-.5in}   \sum_{i=1}^\ell -\log\det\nolimits  \mathbf{Q}_{W,i}   + \text{tr} \left(  (\mathbf{S}_{1,i} + \mu \boldsymbol{\eta} \boldsymbol{\eta}^{\mathrm{T}}) \mathbf{Q}_{W,i} \right) + \lambda_W \left( \alpha_W \|\mathbf{Q}_{W,i} - \mathbf{T}_i \|_1 + \frac{1-\alpha_W}{2} \|\mathbf{Q}_{W,i} - \mathbf{T}_i\|_F^2 \right). \nonumber 
\end{align}

\section{Optimization}

TEASE leverages two major insights for speed: EOF basis and block-diagonal precision matrices in material likelihood, and sparse matrix operations for immaterial likelihood.

Optimizing $\kappa_0$ was numerically unstable in simulation studies. 
For this reason, we fix $\kappa_0$ and focus on other parameters.
One option is to run the procedure for several values of $\kappa_0$ and select the best result.

Algorithm loops through steps in next subsections and terminates when successive changes between $\boldsymbol{\alpha}, \boldsymbol{\Gamma}_1 \boldsymbol{\eta}$, $\mathbf{Q}_W$, and $\mathbf{Q}_V$ are under tolerance $0.05$, or maximum number of iterations is reached.

\label{sec:opt}

\subsubsection{$\boldsymbol{\eta}$ part}

Gradient of material trace
\begin{align}
  \label{eq:materialtrace}
  \hspace{-.5in}
&\frac{1}{t} \sum_{k=1}^t \left( \text{vec}(\boldsymbol{\Gamma}_1^{\mathrm{T}} \tilde{\mathbf{Y}}_k \boldsymbol{\Psi}_{1}) - (\boldsymbol{\Psi}_{1}^{\mathrm{T}} {\mathbf{X}}_k^{\mathrm{T}} \otimes \mathbf{I}_u) \text{vec}(\boldsymbol{\eta})\right)^{\mathrm{T}} \mathbf{Q}_W \left( \text{vec}(\boldsymbol{\Gamma}_1^{\mathrm{T}} \tilde{\mathbf{Y}}_k \boldsymbol{\Psi}_{1}) - (\boldsymbol{\Psi}_{1}^{\mathrm{T}} {\mathbf{X}}_k^{\mathrm{T}} \otimes \mathbf{I}_u) \text{vec}(\boldsymbol{\eta})\right)
\end{align}
with respect to $\text{vec}(\boldsymbol{\eta})$ equals
\begin{equation}
\label{eq:materialtracegradient}
-\frac{2}{t} \sum_{k=1}^t ({\mathbf{X}}_k \boldsymbol{\Psi}_{1} \otimes \mathbf{I}_u) \mathbf{Q}_W \left( \text{vec}(\boldsymbol{\Gamma}_1^{\mathrm{T}} \tilde{\mathbf{Y}}_k \boldsymbol{\Psi}_{1}) - (\boldsymbol{\Psi}_{1}^{\mathrm{T}} {\mathbf{X}}_k^{\mathrm{T}} \otimes \mathbf{I}_u) \text{vec}(\boldsymbol{\eta})\right),
\end{equation}
and gradient of enhanced penalty 
\begin{equation}
\label{eq:enhancedepenalty}
 \mu \sum_{i=1}^\ell \text{vec}(\boldsymbol{\eta})^{\mathrm{T}} (\mathbf{I}_p \otimes \mathbf{Q}_{W,i}) \text{vec}(\boldsymbol{\eta}) \\
\end{equation}
with respect to $\text{vec}(\boldsymbol{\eta})$ equals
\begin{equation}
\label{eq:enhancedepenaltygradient}
2 \mu \sum_{i=1}^{\ell}  (\mathbf{I}_p \otimes \mathbf{Q}_{W,i}) \text{vec}(\boldsymbol{\eta}),
\end{equation}
leading to a sparse linear system
\begin{align}
&\left( \frac{1}{t} \sum_{k=1}^t ({\mathbf{X}}_k \boldsymbol{\Psi}_{1} \otimes \mathbf{I}_u) \mathbf{Q}_W (\boldsymbol{\Psi}_{1}^{\mathrm{T}} {\mathbf{X}}_k^{\mathrm{T}} \otimes \mathbf{I}_u)  + \mu \sum_{i=1}^{\ell}  (\mathbf{I}_p \otimes \mathbf{Q}_{W,i}) \right) \text{vec}(\boldsymbol{\eta}) \\ \nonumber
&=
\frac{1}{t} \sum_{k=1}^t ({\mathbf{X}}_k \boldsymbol{\Psi}_{1} \otimes \mathbf{I}_u) \mathbf{Q}_W \text{vec}(\boldsymbol{\Gamma}_1^{\mathrm{T}} \tilde{\mathbf{Y}}_k \boldsymbol{\Psi}_{1}).
\end{align}

\subsubsection{$\mathbf{Q}$ material part}

Due to orthogonality, minimization separates into independent targeted graphical elastic net \citep{kovács2021graphicalelasticnettarget} problems ($i=1,\dots,\ell$):
\begin{equation}
\label{eq:targetedelasticnet_material}
- \log \det \mathbf{Q}_{W,i} + \text{tr}( \mathbf{S}^*_{1,i}   \mathbf{Q}_{W,i}) + \lambda_W \left( \alpha_W \|\mathbf{Q}_{W,i} - \mathbf{T}_i \|_1 + \frac{1-\alpha_W}{2} \|\mathbf{Q}_{W,i} - \mathbf{T}_i \|_F^2 \right).
\end{equation}
\eqref{eq:targetedelasticnet_material} shrinks $\mathbf{Q}_{W,i}$ towards target $\mathbf{T}_i$, equal to $\mathbf{I}_{u}$ times reciprocal of $i$th EOF eigenvalue.
Tuning parameter $\alpha_W \in [0,1]$ interpolates between graphical ridge \citep{VANWIERINGEN2020104621,VANWIERINGEN2016284,Kuismin03072017} and graphical lasso \citep{yuanlin, banerjee_model_2008, friedman_sparse_2008, QUIC}, 
and $\lambda_W >0$ controls overall regularization strength.
Targeted fused ridge penalty \citep{anders2020} is another option for targeted precision matrix estimation.
In special cases, \eqref{eq:targetedelasticnet_material} has closed-form solution. 
Unregularized solution ($\lambda_W=0$) is $(\mathbf{S}_{1,i}^*)^{-1}$.
Graphical ridge ($\alpha_W = 0$) solution is 
$\dfrac{1}{\lambda_W} \left(\sqrt{\lambda_W \mathbf{I}_{u} + (\mathbf{S}^*_{1,i}-\mathbf{T}_i)^2/4}  - (\mathbf{S}^*_{1,i}-\mathbf{T}_i)/2\right)$ \citep{VANWIERINGEN2016284}; the matrix square root follows readily from an eigendecomposition of $\mathbf{S}^*_{1,i} - \mathbf{T}_i$.
A recent breakthrough in graphical lasso regularization \citep{eftekhari2026regularizationparametersparseprecision} could be adapted for targeted graphical elastic net.

\subsubsection{$\mathbf{Q}$ immaterial part}

Adding $\delta \mathbf{I}_{n(r-u)}$ for stability, $0 < \delta \ll 1$, followed by Sherman-Morrison-Woodbury matrix identity, \eqref{eq:immateriallogdet} equals $n (r-u) \log \delta$ plus
\begin{equation}
  \label{eq:logdettrick}
  \hspace{-.6in}
-\left(  \sum_{h=1}^{m_\ell} n_h \log \det (\mathbf{Q}_{V,h}) + 2 (r-u) \log \det (\mathbf{B}_h)  \right) + \log \det \left(  \delta^{-1} (\boldsymbol{\Phi}_0^{\mathrm{T}} \boldsymbol{\Phi}_0 \otimes \mathbf{I}_{r-u} ) + \bigoplus_{h=1}^{m_\ell} \mathbf{Q}_{V,h} \otimes (\mathbf{B}_h \mathbf{B}_h^{\mathrm{T}})  \right).
\end{equation} 
Since multiresolution bases are different sizes $\{n_h\}_{h=1}^{m_\ell}$, last log determinant in \eqref{eq:logdettrick} does not simplify but is highly-structured
and quickly computed via sparse Cholesky decomposition.
Nonetheless, it is a bottleneck; 
we leverage a difference-of-convex \citep{dc} matrix identity:
\begin{align}
  \label{eq:linearize}
& \log \det  \left( \delta^{-1} (\boldsymbol{\Phi}_0^{\mathrm{T}} \boldsymbol{\Phi}_0 \otimes  \mathbf{I}_{r-u}   )  + \bigoplus_{h=1}^{m_\ell} \mathbf{Q}_{V,h} \otimes (\mathbf{B}_h \mathbf{B}_h^{\mathrm{T}})  \right) \le \log \det \boldsymbol{\Upsilon} \\ \nonumber
& + \text{tr}\left(\boldsymbol{\Upsilon}^{-1} \left( \bigoplus_{h=1}^{m_\ell} \mathbf{Q}_{V,h} \otimes (\mathbf{B}_h \mathbf{B}_h^{\mathrm{T}}) - \mathbf{Q}_{V,h}^* \otimes (\mathbf{B}_h \mathbf{B}_h)^{\mathrm{T}}) \right)\right),
\end{align}
$\boldsymbol{\Upsilon} = \delta^{-1} ( \boldsymbol{\Phi}_0^{\mathrm{T}} \boldsymbol{\Phi}_0 \otimes \mathbf{I}_{r-u} )   + \bigoplus_{h=1}^{m_\ell} \mathbf{Q}_{V,h}^* \otimes (\mathbf{B}_h \mathbf{B}_h)^{\mathrm{T}}$, $\{\mathbf{Q}_{V,h}^*\}_{h=1}^{m_\ell}$ are previous guesses.
Summing diagonal elements produces quadratic forms to leverage \citet[11.17]{matrixhandbook}:
\begin{equation}
  \label{eq:immaterialtrace2}
  \hspace{-.3in}
\sum_{i=1}^{m (r-u)} \boldsymbol{\nabla}_i^{\mathrm{T}} \left(  \bigoplus_{h=1}^{m_\ell} \mathbf{Q}_{V,h} \otimes (\mathbf{B}_h \mathbf{B}_h^{\mathrm{T}})  \right) \mathbf{e}_i
= \sum_{h=1}^{m_\ell}  \text{tr}\left( \left( \sum_{i=1}^{m(r-u)} \text{mat}(\boldsymbol{\nabla}_{i,h})^{\mathrm{T}} \mathbf{B}_h \mathbf{B}_h^{\mathrm{T}} \ \text{mat}(\mathbf{e}_{i,h})  \right) \mathbf{Q}_{V,h} \right)
\end{equation}
with $\boldsymbol{\nabla}_i = \mathbf{e}_i^{\mathrm{T}} \boldsymbol{\Upsilon}^{-1}, \mathbf{e}_i$ one-hot vector, and $\boldsymbol{\nabla}_{i,h}, \mathbf{e}_{i,h}$ extract multiresolution level $h$ indices.
Sparse linear solves $\mathbf{e}_i^{\mathrm{T}} \boldsymbol{\Upsilon}^{-1} $ and block diagonal Kronecker mat-vec multiplications are rapid.
Similarly, rewrite immaterial trace:
\begin{equation}
\label{eq:immaterialtracerewrite}
\text{tr}(\mathbf{S}_0 \mathbf{Q}_V) = \sum_{h=1}^{m_\ell}  \text{tr}\left( \left( \frac{1}{t} \sum_{k=1}^t\text{mat}(v_{k,h})^{\mathrm{T}}\mathbf{B}_h \mathbf{B}_h^{\mathrm{T}} \text{mat}(v_{k,h}) \right) \mathbf{Q}_{V,h} \right),
\end{equation} $v_{k,h} $ extracts level $h$ elements from $\text{vec}(\boldsymbol{\Gamma}_0^{\mathrm{T}} \tilde{\mathbf{Y}}_k \boldsymbol{\Psi}_0)$.
Combining \eqref{eq:immaterialtracerewrite}, \eqref{eq:immaterialtrace2}, and \eqref{eq:logdettrick} forms independent sparse covariance estimation problems \citep{bien}, written with $\boldsymbol{\Sigma}_{V,h} = \mathbf{Q}_{V,h}^{-1}$:
\begin{equation}
\label{eq:spcov_immaterial}
\log \det \boldsymbol{\Sigma}_{V,h} + \text{tr}( \mathbf{S}_{0,h}   \boldsymbol{\Sigma}_{V,h}^{-1}) +  \|\boldsymbol{\Lambda}_V  \circ \boldsymbol{\Sigma}_{V,h} \|_1 \quad (h=1,\dots,m_\ell), 
\end{equation}
$\mathbf{S}_{0,h} = \dfrac{1}{n_h} \left( \left( \displaystyle \sum_{i=1}^{\ell u} \text{mat}(\boldsymbol{\nabla}_{i,h})^{\mathrm{T}} \mathbf{B}_h \mathbf{B}_h^{\mathrm{T}} \ \text{mat}(\mathbf{e}_{i,h})  \right) 
+ \left( \dfrac{1}{t} \displaystyle \sum_{k=1}^t\text{mat}(v_{k,h})^{\mathrm{T}}\mathbf{B}_h \mathbf{B}_h^{\mathrm{T}} \text{mat}(v_{k,h}) \right)  \right)$, $\circ$ is entrywise product, and $\boldsymbol{\Lambda}_V$ zero except for positive diagonal entries.
Regularization is performed on covariances rather than precisions due to overestimated variances in the latter case. 
Similar behavior was noted by \citet{caringi2026fixedrankcokrigingmodel}.
Crucially, constructing $\boldsymbol{\Lambda}_V = \lambda_V \mathbf{I}_{r-u}$ with zero off-diagonal penalty prevents degenerate diagonal solutions. 
See \citet{vanWieringen2024} for the ridge regularized version of \eqref{eq:spcov_immaterial}.

 For massive meshes, linearized trace dominates computation time and becomes a constraint. 
 Removing it is no sweat with the Hutchinson trace estimator \citep{Hutchinson01011990,anitescu2012,stein2013,Geoga02042020}.
 Replace one-hot vectors $\{\mathbf{e}_i\}_{i=1}^{\ell u}$ by a small number of independent Rademacher vectors $\{\mathbf{g}_i\}_{i=1}^G$ distributed as $-1$ or $1$ with equal probability, and divide by $G$.
Hutchinson Kronecker trace estimates require care \citep{meyer2026}.
Variance reduction can be computed at little extra cost \citep[Hutch++]{meyer2021,persson2022}.
See \cite{chen2016,skorski2020modernanalysishutchinsonstrace} for accuracy results on Hutchinson estimation.

\subsubsection{Stiefel optimization}
\texttt{Manopt.jl} \citep{Bergmann2022} is leveraged to minimize
\begin{align}  
  \label{eq:stiefel}
  & \frac{1}{t} \displaystyle \sum_{k=1}^t \text{vec}(\boldsymbol{\Gamma}_1^{\mathrm{T}} \tilde{\mathbf{Y}}_k \boldsymbol{\Phi}_1 -\boldsymbol{\eta} {\mathbf{X}}_k \boldsymbol{\Phi}_{1})^{\mathrm{T}} \mathbf{Q}_W \text{vec}(\boldsymbol{\Gamma}_1^{\mathrm{T}} \tilde{\mathbf{Y}}_k \boldsymbol{\Phi}_1 - \boldsymbol{\eta} {\mathbf{X}}_k \boldsymbol{\Phi}_{1}) \\ \nonumber
  + & \frac{1}{t} \displaystyle \sum_{k=1}^t \text{vec}(\boldsymbol{\Gamma}_0^{\mathrm{T}} \tilde{\mathbf{Y}}_k \boldsymbol{\Psi}_0)^{\mathrm{T}} \mathbf{Q}_V \text{vec}(\boldsymbol{\Gamma}_0^{\mathrm{T}} \tilde{\mathbf{Y}}_k \boldsymbol{\Psi}_0)
\end{align}
over $\boldsymbol{\Gamma} = [\boldsymbol{\Gamma}_1 \ \boldsymbol{\Gamma}_0] \in \mathbb{R}^{r,r}$ subject to $\boldsymbol{\Gamma}^{\mathrm{T}} \boldsymbol{\Gamma} = \mathbf{I}_r$.
This step could be accelerated with a specialized linear solver \citep{li2020}.

\subsubsection{$\boldsymbol{\alpha}$ part}
Intercept $\boldsymbol{\alpha}$ is estimated from residuals:
\begin{equation}
\hat{\boldsymbol{\alpha}} = \frac{1}{tn} \sum_{k=1}^t \sum_{i=1}^n \mathbf{Y}_k(\mathbf{s}_i) - \hat{\boldsymbol{\Gamma}}_1 \hat{\boldsymbol{\eta}} \mathbf{X}_k(\mathbf{s}_i).
\end{equation}

\section{Data Analysis}
\label{sec:data}

TEASE exhibitions are laid out on two real data examples.
First, an analysis of anomaly fields from NCAR's Community Atmosphere Model \citep{gmd-8-2829-2015}, previously studied in \cite{krock2023mbgl}.
Here, response variables are \texttt{TMQ}, \texttt{U10}, \texttt{SWCF}, \texttt{AODVIS}, \texttt{FLNTC}, \texttt{PBLH}, \texttt{PS}, and \texttt{PRECT}; the remaining \texttt{Flux}, \texttt{Cloud}, and \texttt{Aerosol} variables (28 total) are predictors.

Second are gridMET daily high-resolution weather fields in contiguous U.S.\ \citep{Abatzoglou2013}.
Response variables are minimum and maximum relative humidity, specific humidity, and surface downwelling shortwave flux in air; predictor variables are longitude, latitude, elevation, precipitation, minimum and maximum air temperature, 10m wind speed and direction, burning index, energy release component, dead fuel moisture (1, 10, 100, 1000 hours), potential evapotranspiration (grass and alfalfa),  and mean vapor pressure deficit.

For simplicity, $\alpha_W = 1/3$ is chosen to equally weight $\ell_1$ and $\ell_2$ norms in targeted graphical elastic net penalty. 
SAR parameter $\kappa_0^2$ is set to default value 0.05. 
Instead of an expensive cross-validation search for envelope dimension $u$ and hyperparameters $\mu, \lambda_W, \lambda_V$, we select the combination that 
minimizes energy distance \citep{MR3055745} between TEASE simulations and observations.
Handling $\lambda_V$ is of chief importance.
In our data examples, $\mu$ and $\lambda_W$ are set to zero while searching for the optimal $u$ and $\lambda_V$ over a grid of values.
Then, $\mu$ and $\lambda_W$ are chosen with $u,\lambda_V$ fixed.
See \cite{HWAH2024} for sliced energy distance in big data settings.

\subsection{CAM}
Data are standardized before analysis and recorded at $n=48602$ locations over Earth in $t=343$ realizations.
Four levels of Wendland basis functions total $m = 858$, and $\ell= 888$ EOFs explain 95\% of response variance.

\begin{figure}[h!]
    \centering
    \includegraphics[width=2in,height=1in]{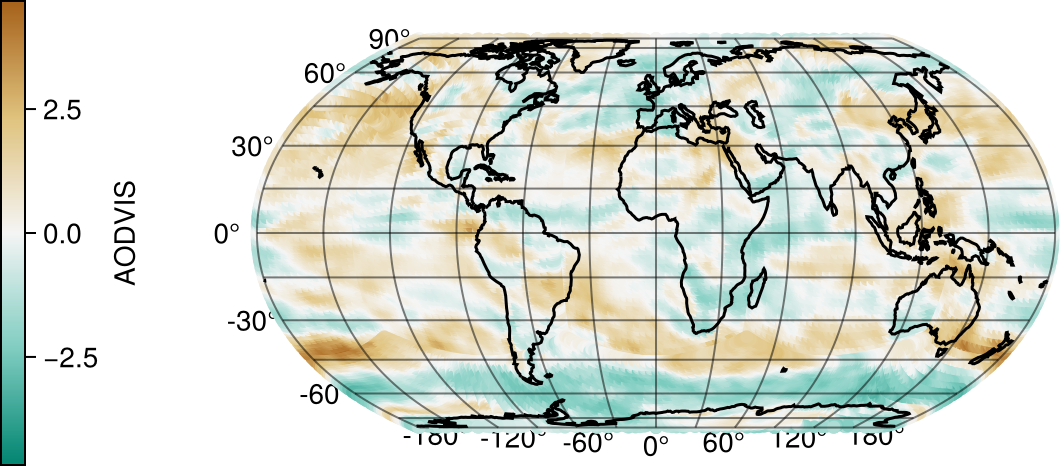}
    \includegraphics[width=2in,height=1in]{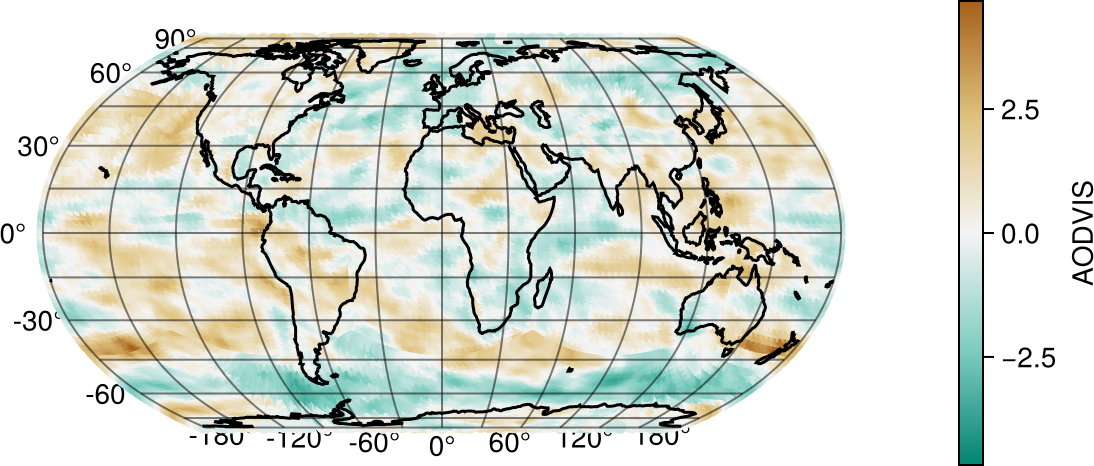}

    \includegraphics[width=2in,height=1in]{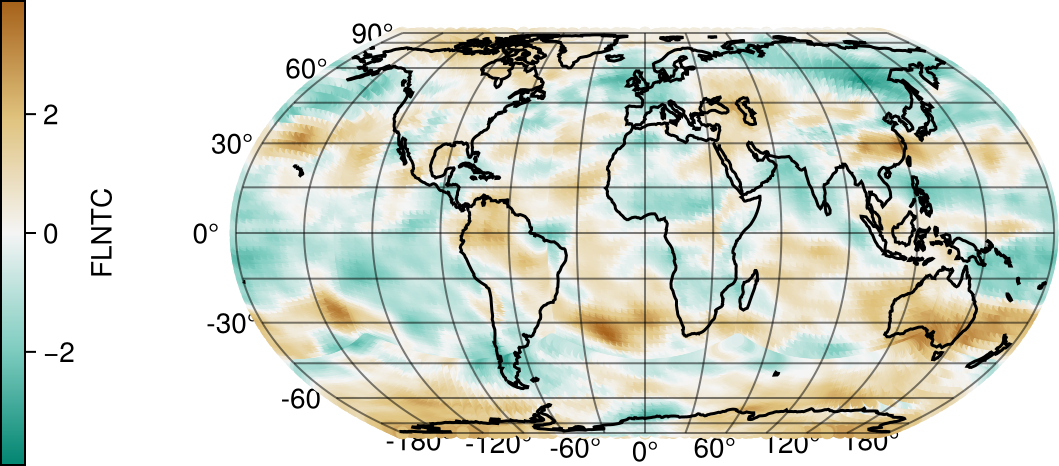}
    \includegraphics[width=2in,height=1in]{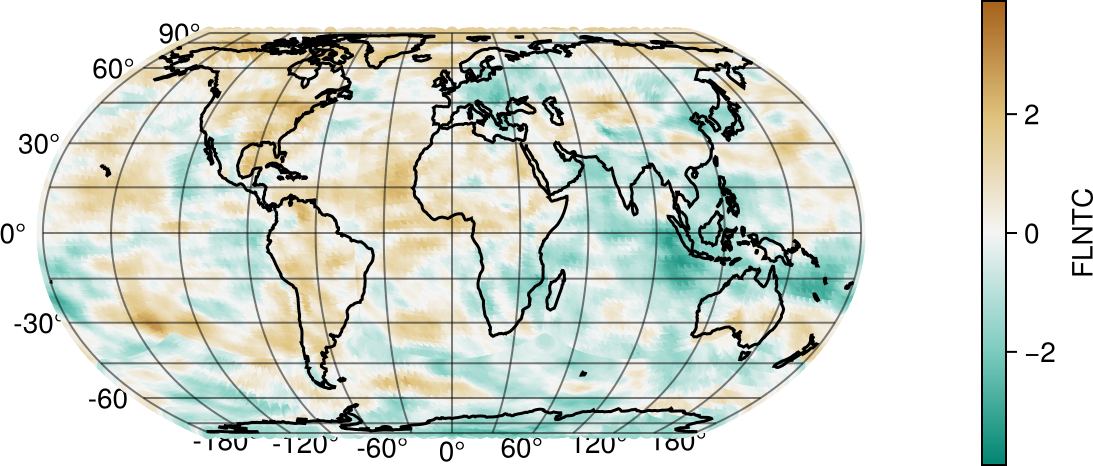}

        \includegraphics[width=2in,height=1in]{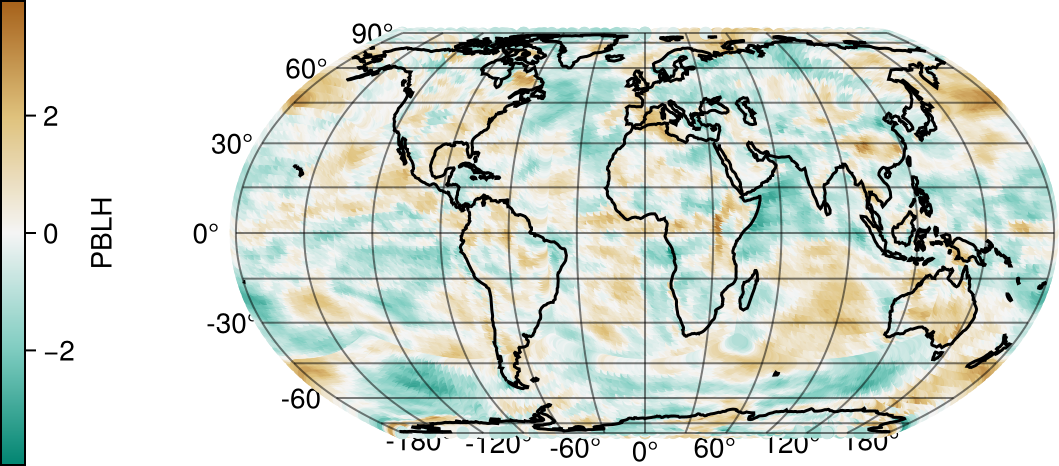}
    \includegraphics[width=2in,height=1in]{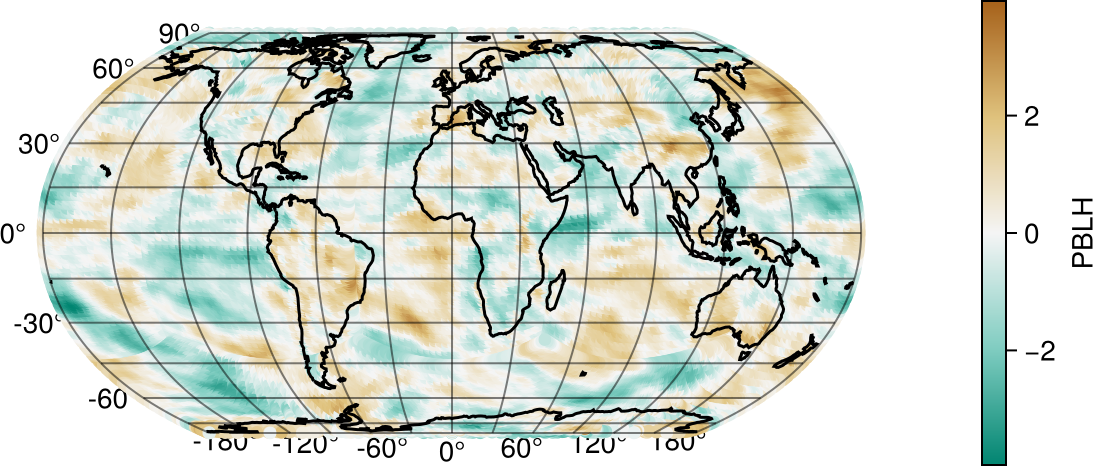}

        \includegraphics[width=2in,height=1in]{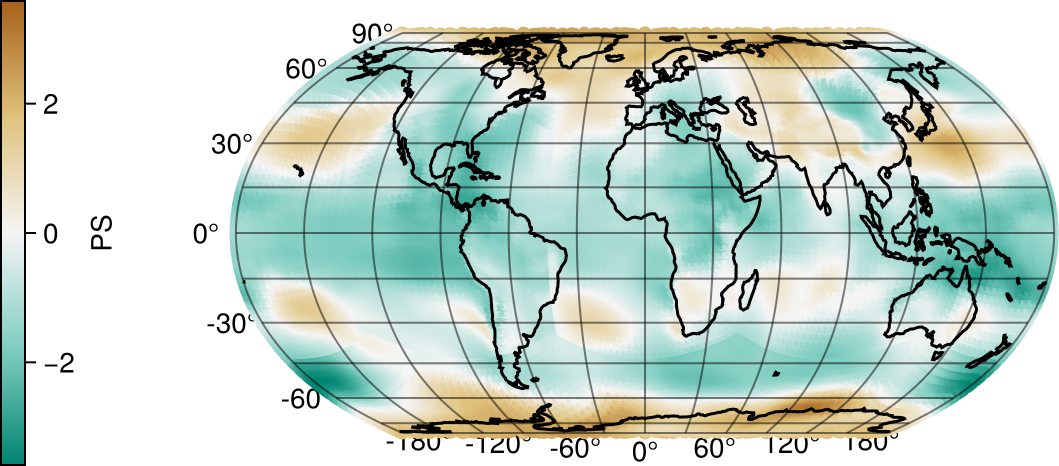}
    \includegraphics[width=2in,height=1in]{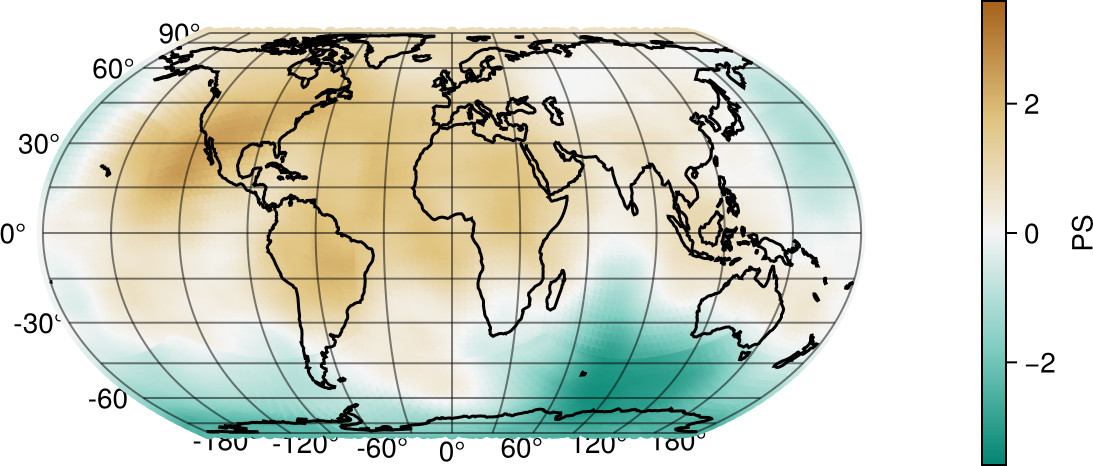}

        \includegraphics[width=2in,height=1in]{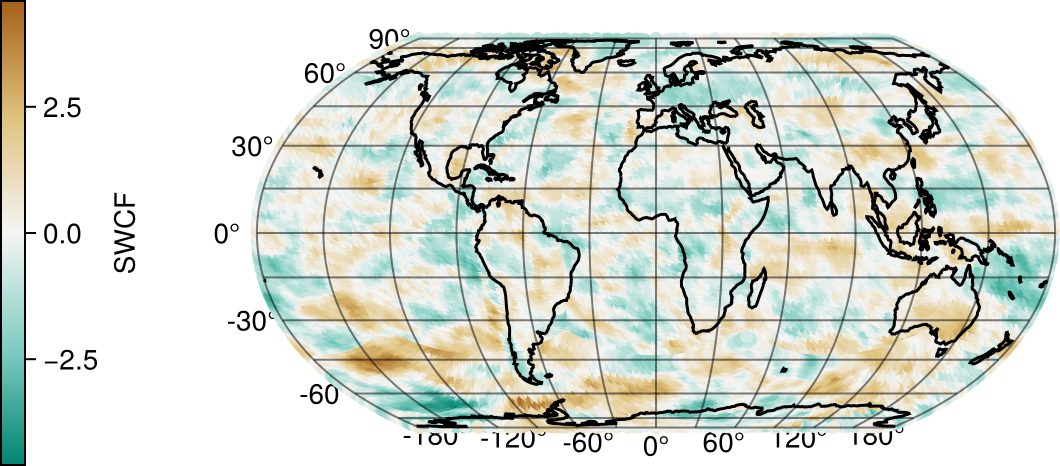}
    \includegraphics[width=2in,height=1in]{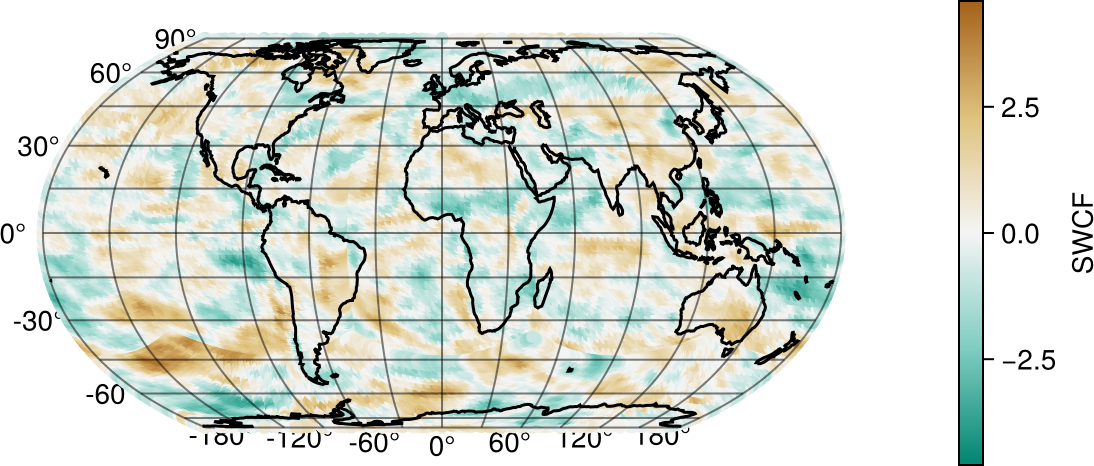}

        \includegraphics[width=2in,height=1in]{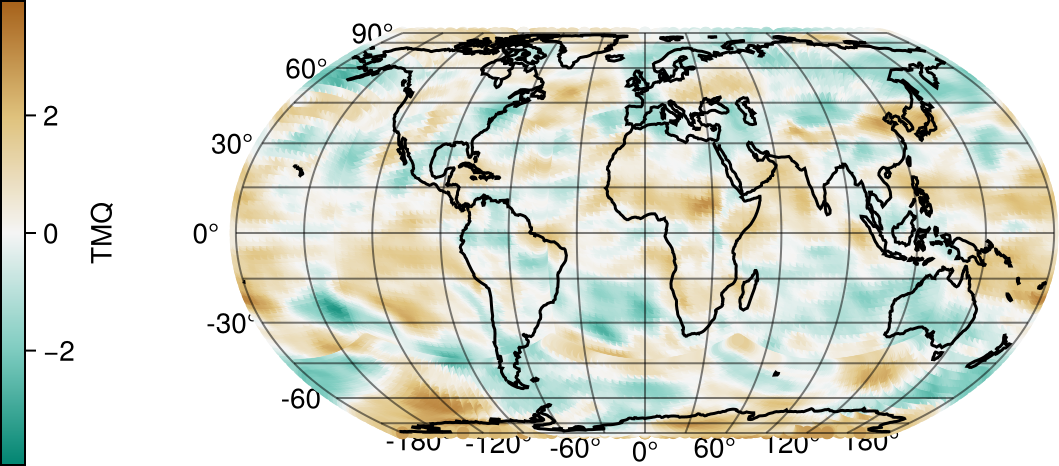}
    \includegraphics[width=2in,height=1in]{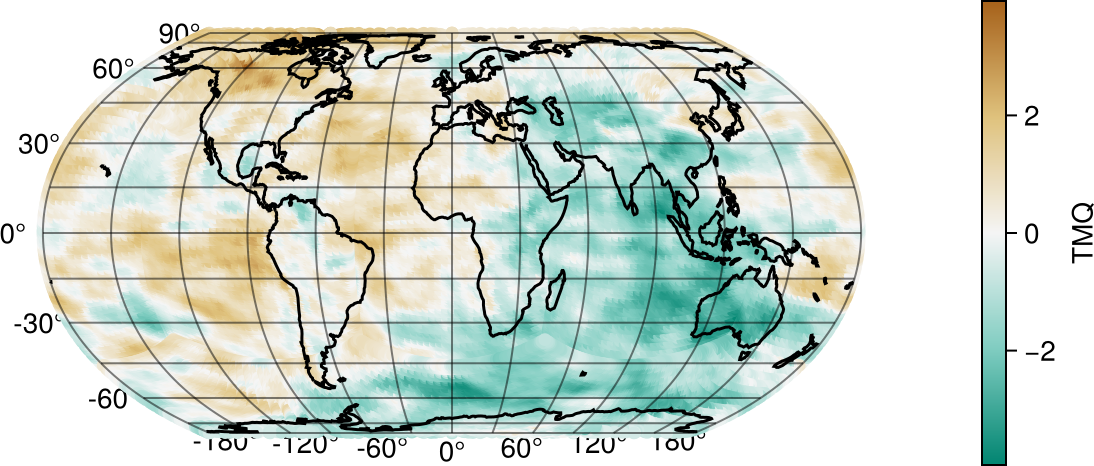}

        \includegraphics[width=2in,height=1in]{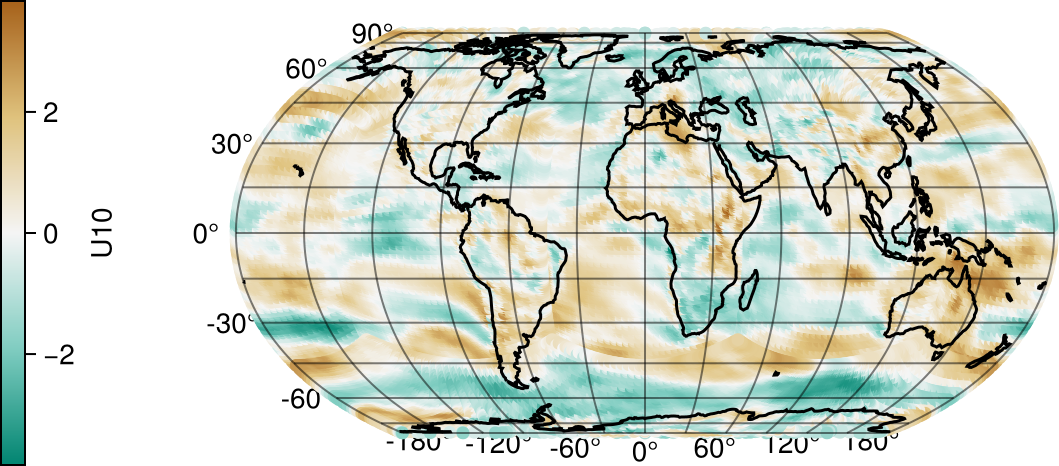}
    \includegraphics[width=2in,height=1in]{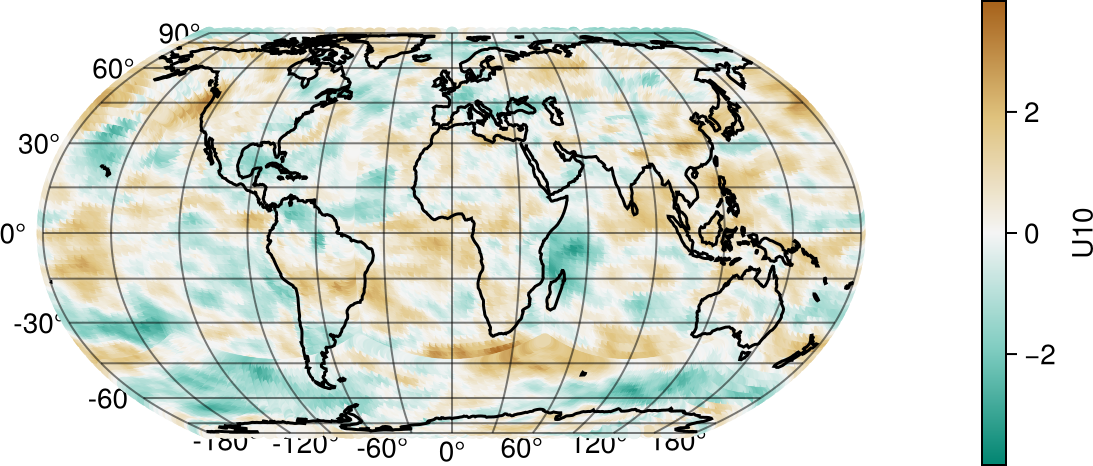}

        \includegraphics[width=2in,height=1in]{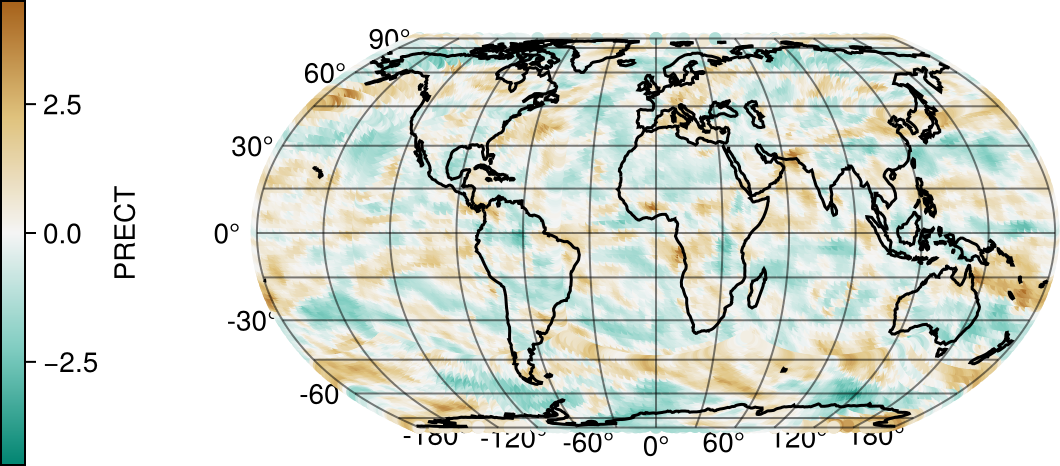}
    \includegraphics[width=2in,height=1in]{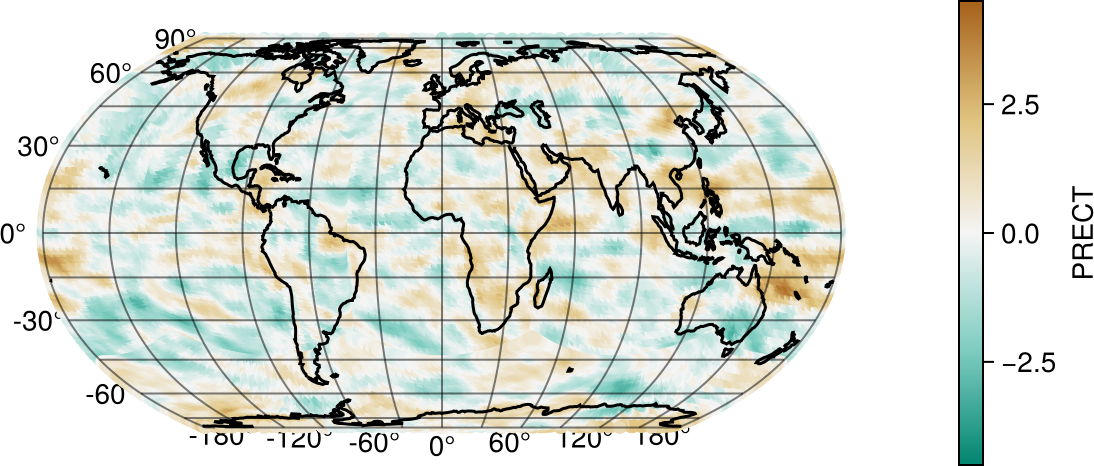}

    \caption{Column 1: CAM data. Column 2: TEASE simulations.}
    \label{fig:datavssims_TAMCDS}
\end{figure}

Perhaps counterintuitively, smoothness of \texttt{PS} comes from immaterial Wendland basis, not EOF basis.
Thus, current TEASE incarnation may be most suitable when response variables have similar smoothness.
Moreover, immaterial basis dominates standard error plots, hence different visual appearance of \texttt{PS} than when solely leveraging EOFs \citep{krock2023mbgl}.

\begin{figure}[h!]
    \centering
    \includegraphics[width=2in,height=1in]{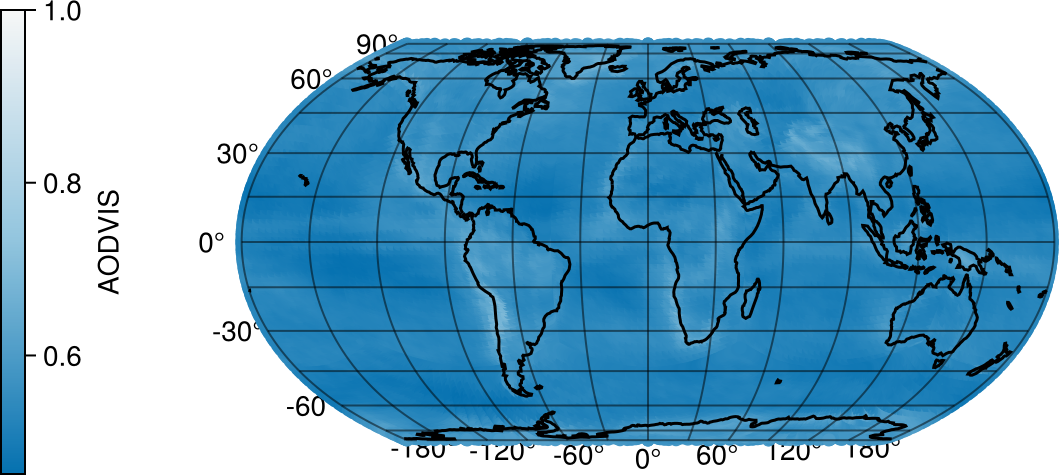}
    \includegraphics[width=2in,height=1in]{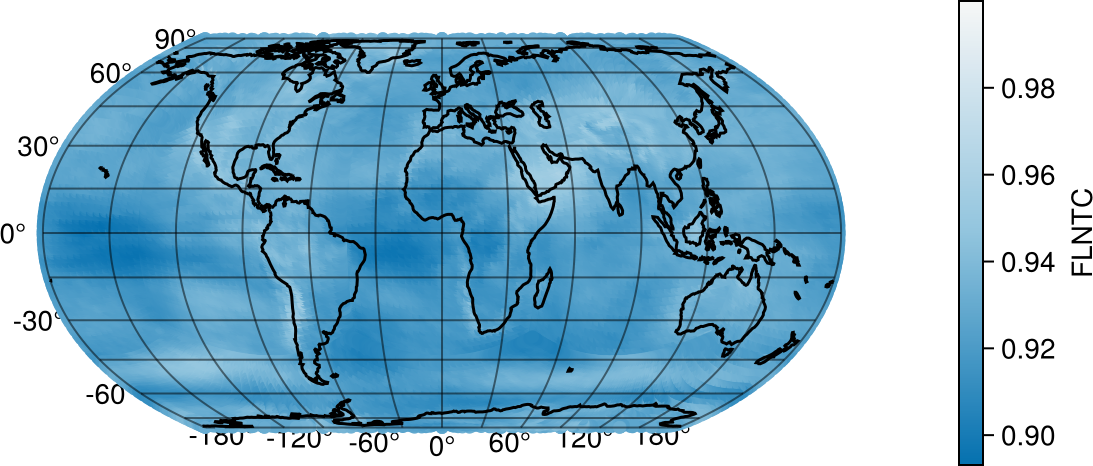}

    \includegraphics[width=2in,height=1in]{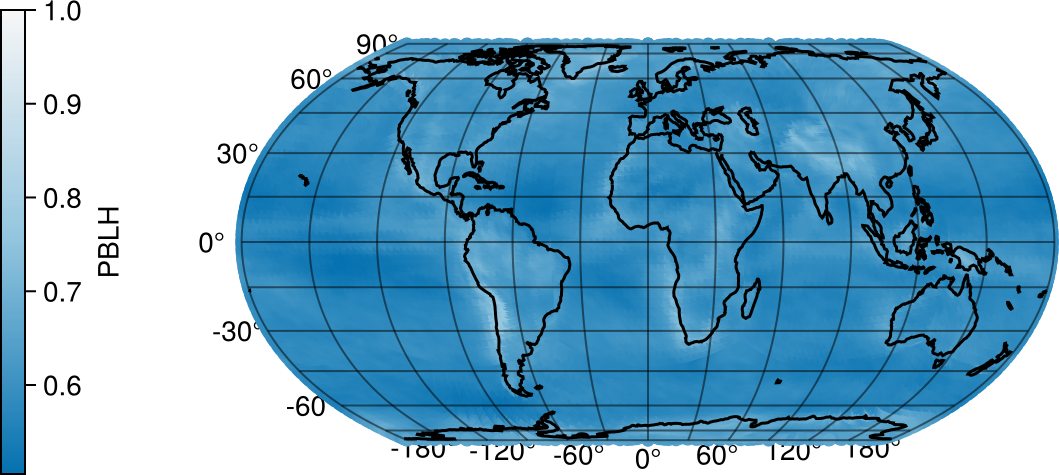}
    \includegraphics[width=2in,height=1in]{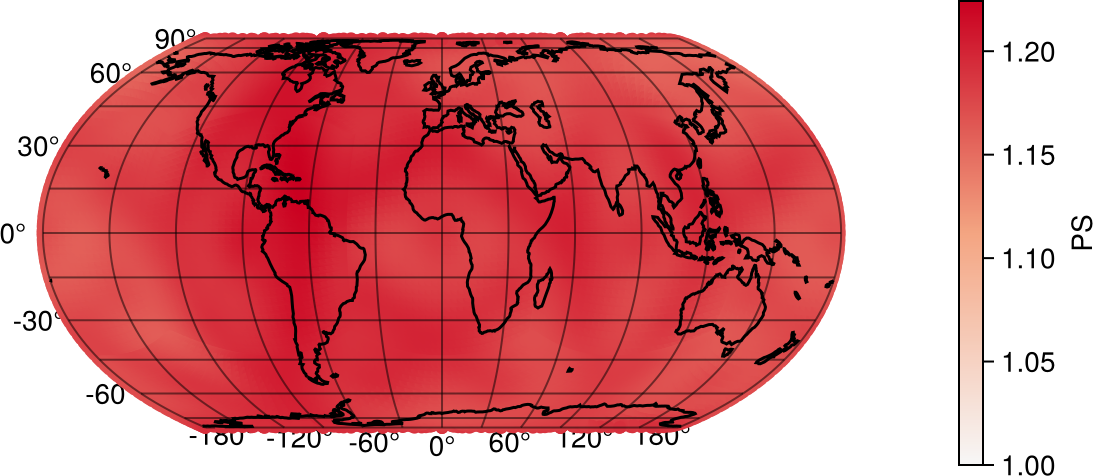}

    \includegraphics[width=2in,height=1in]{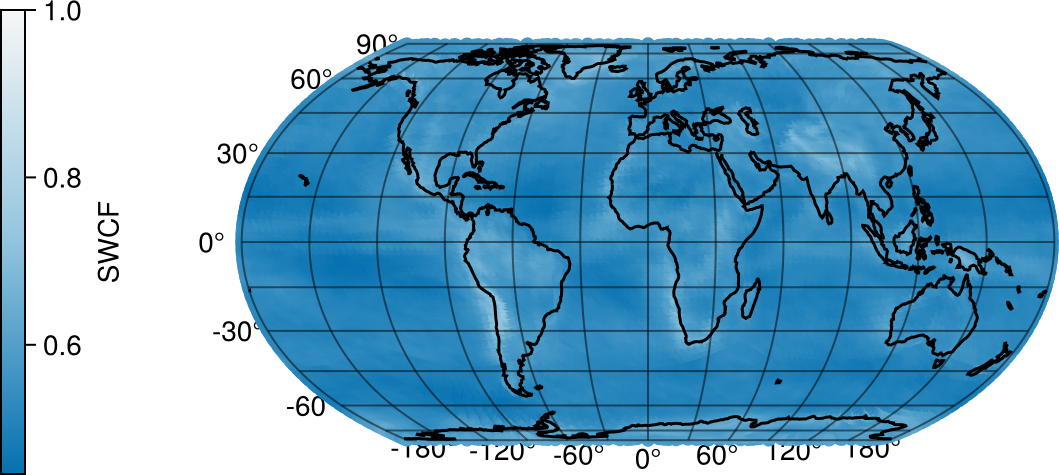}
    \includegraphics[width=2in,height=1in]{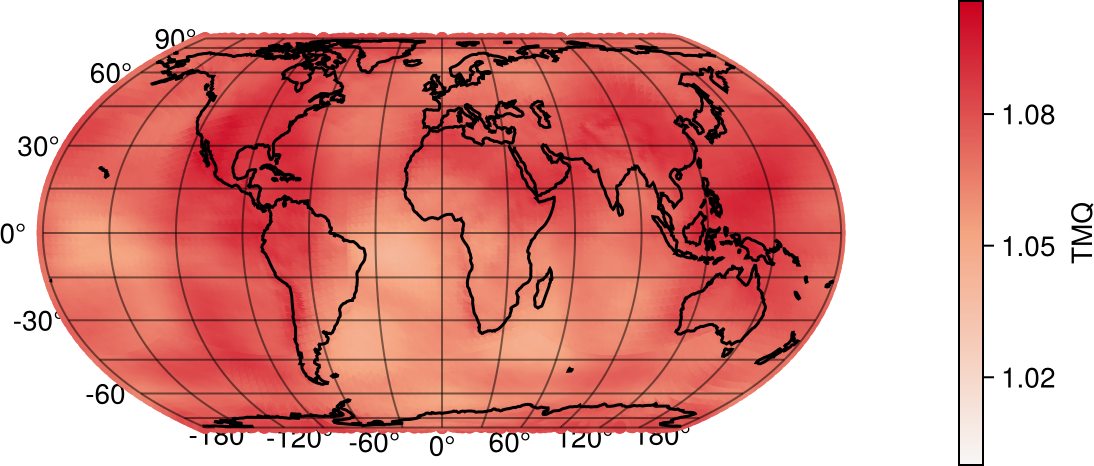}

    \includegraphics[width=2in,height=1in]{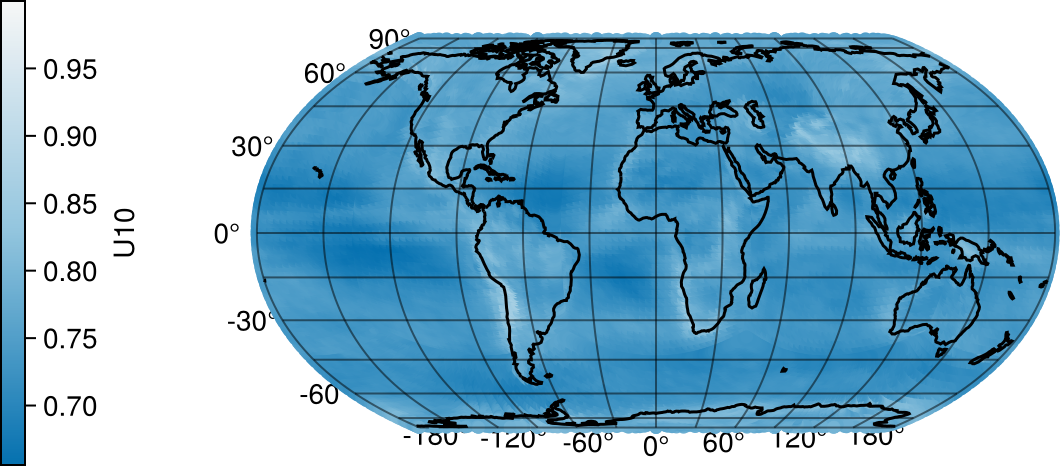}
    \includegraphics[width=2in,height=1in]{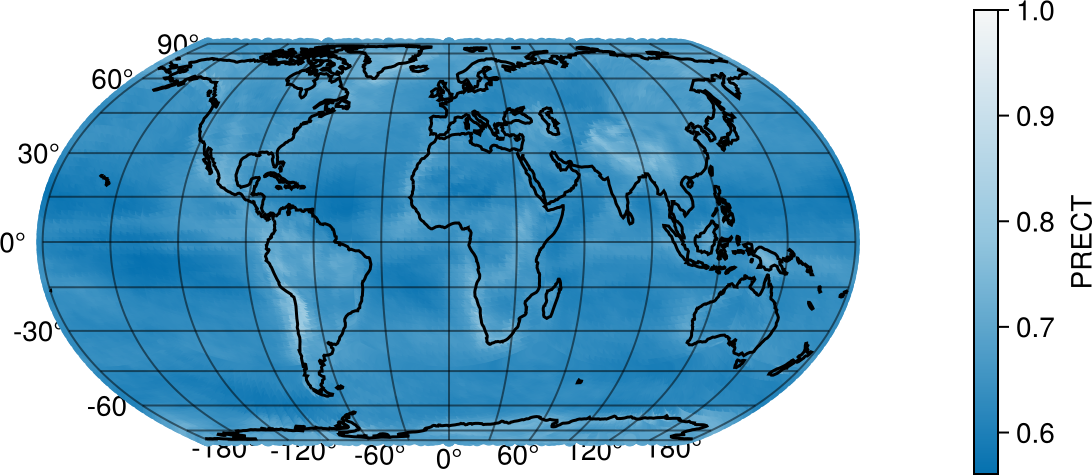}

    \caption{TEASE standard deviations. (Scale white at 1).}
    \label{fig:mean_metdata}
\end{figure}

\begin{figure}[h!]
    \centering
   
    \includegraphics[width=2in,height=1in]{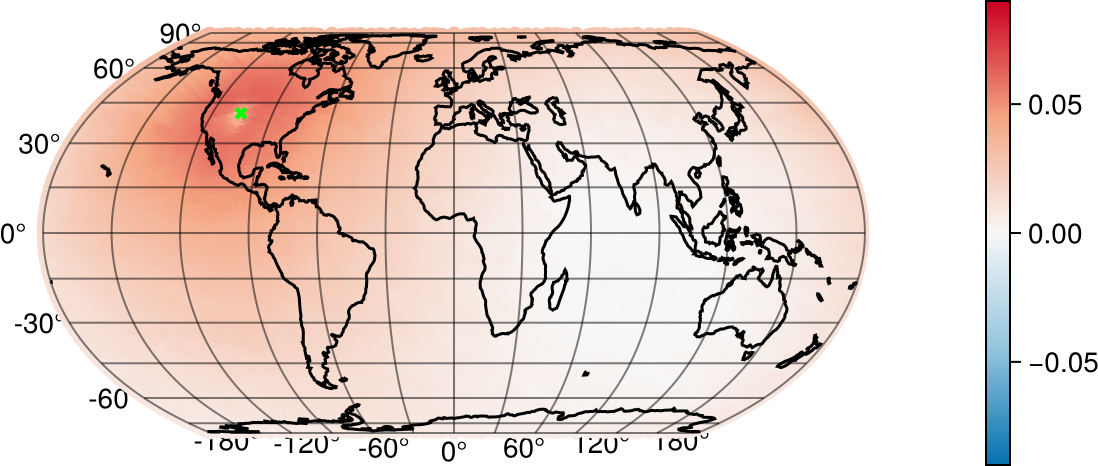}
    \includegraphics[width=2in,height=1in]{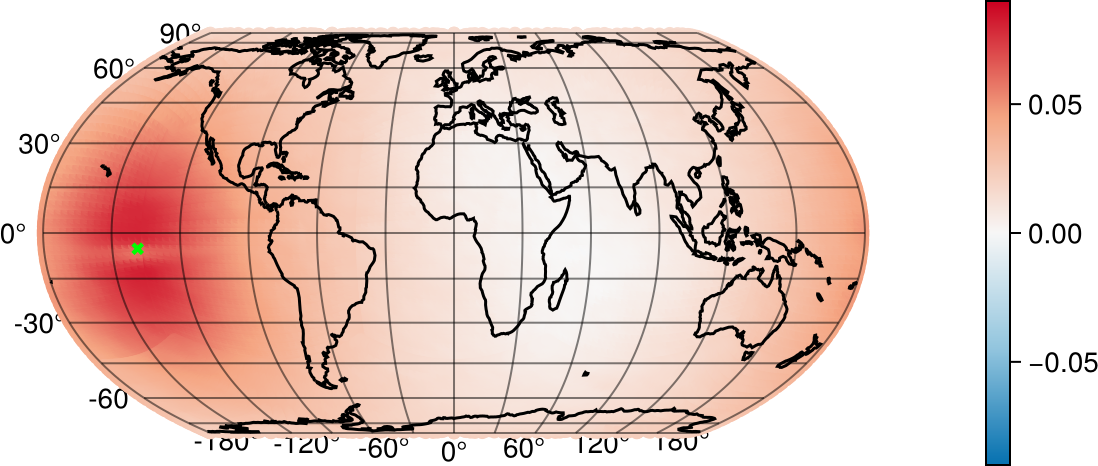}

    \caption{TEASE cross-correlation between \texttt{PRECT} at green $\mathbf{x}$ and \texttt{PS} elsewhere.}
    \label{fig:datavssims}
\end{figure}

\subsection{gridMET}
Gridded data $(n=483879)$ of daily estimates in January 2025 are reported.
Five layer resolution Wendland basis consists of $m = 2080$ functions, and $\ell = 61$ EOFs explain 99\% of response variance.
Unlike previous dataset, no preprocessing via standardization occurs.
However, we model logarithm of the response variables.

\begin{figure}[h!]
    \centering
    \includegraphics[width=2in,height=1in]{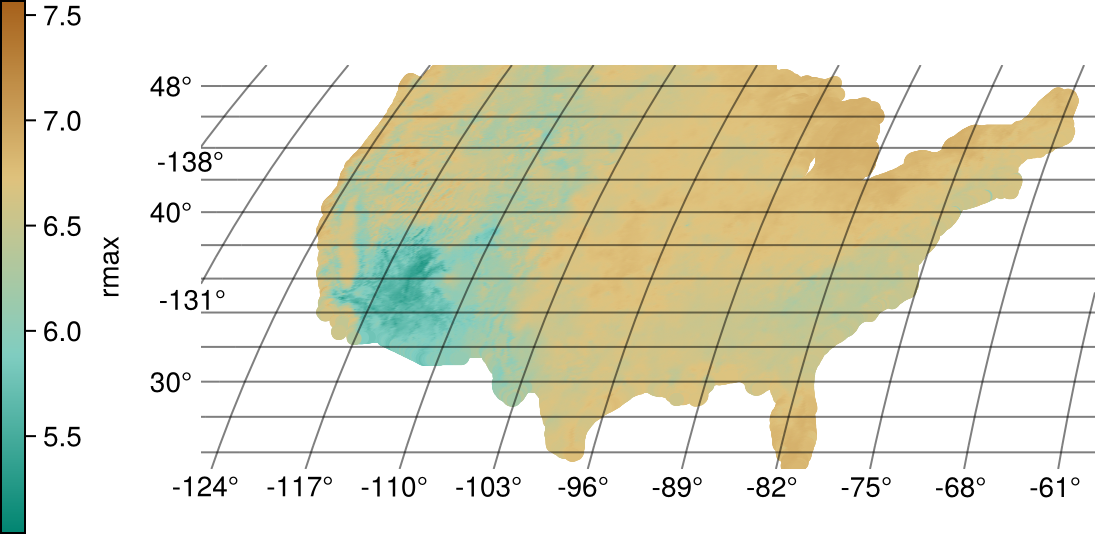}
    \includegraphics[width=2in,height=1in]{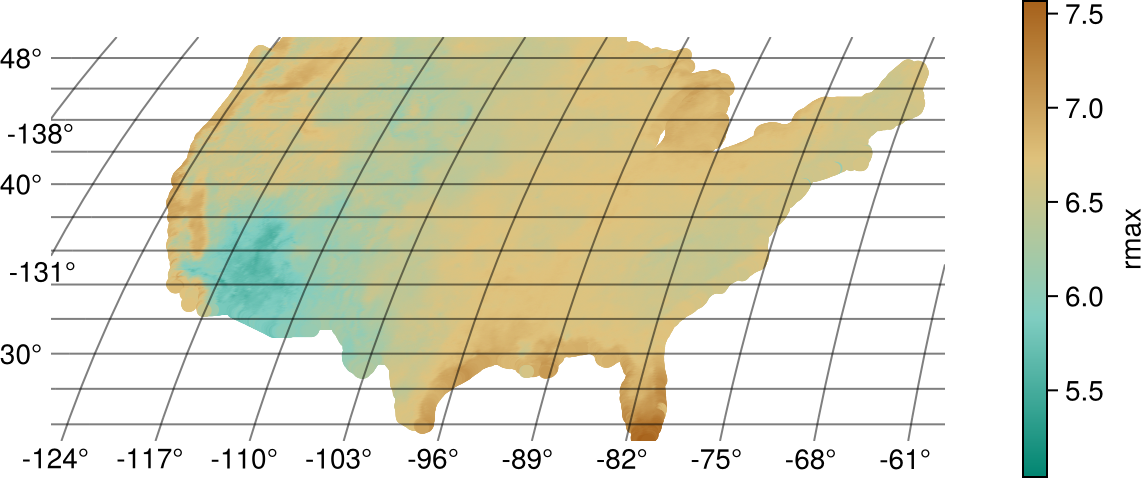}

    \includegraphics[width=2in,height=1in]{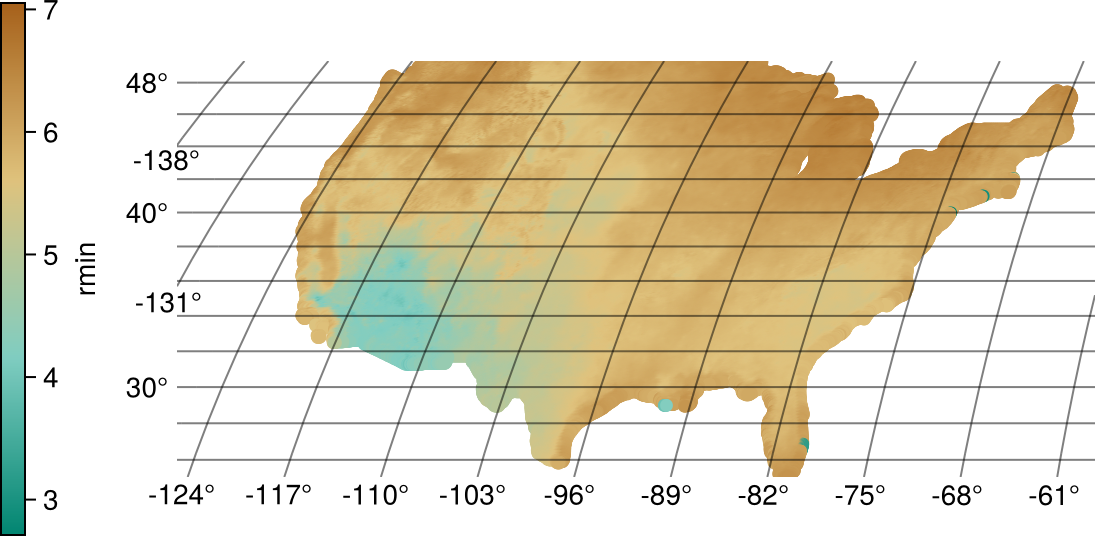}
    \includegraphics[width=2in,height=1in]{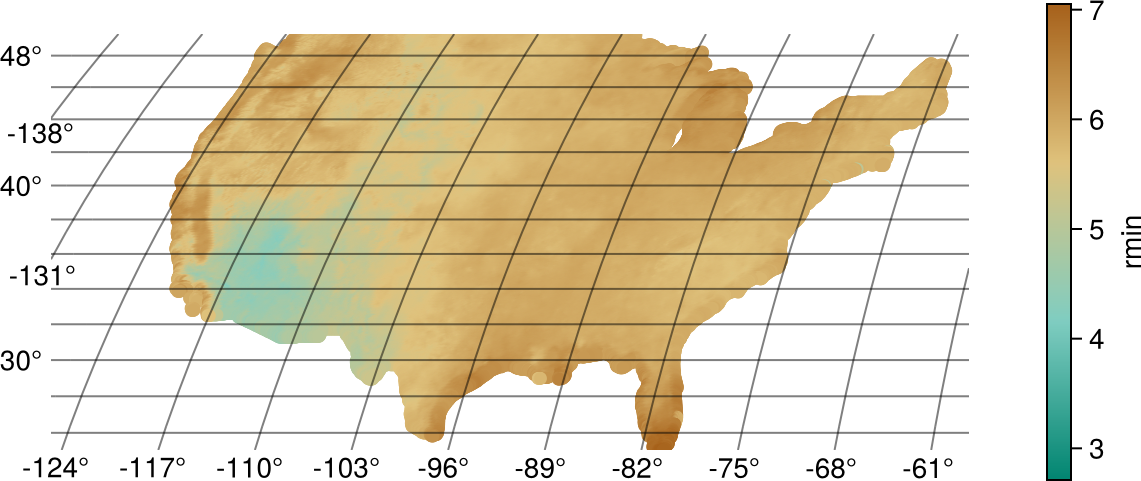}

    \includegraphics[width=2in,height=1in]{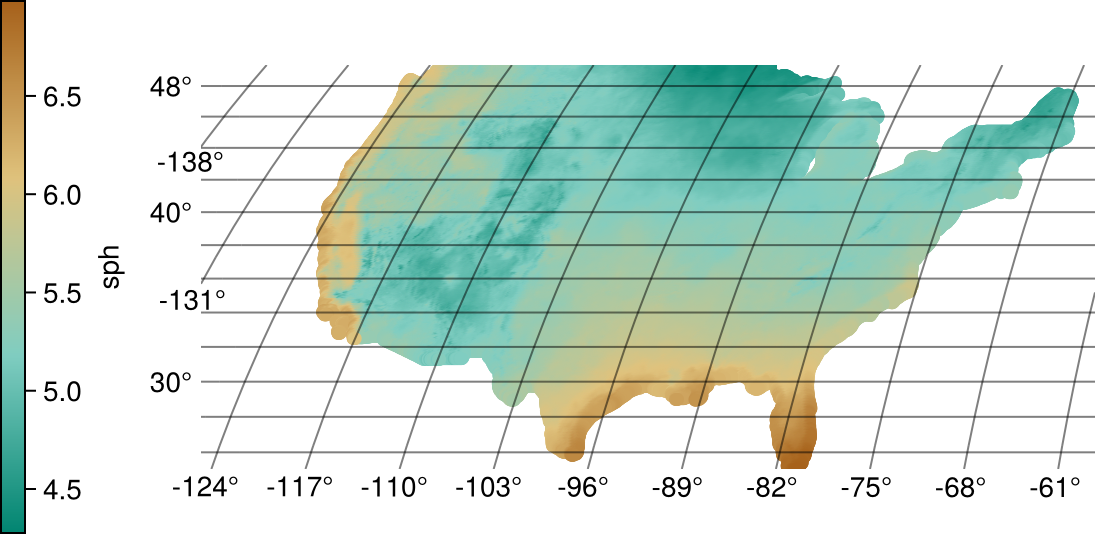}
    \includegraphics[width=2in,height=1in]{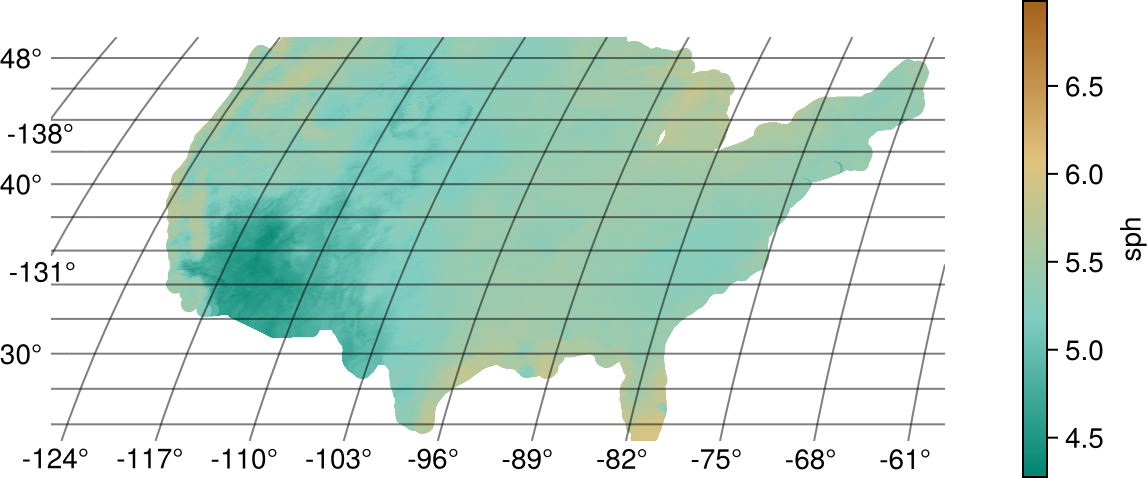}

    \includegraphics[width=2in,height=1in]{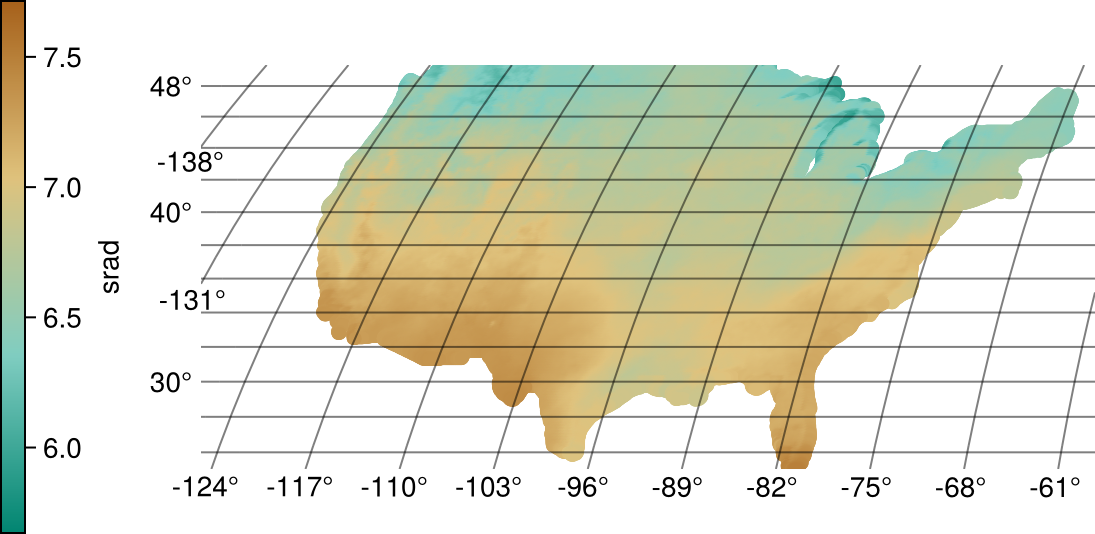}
    \includegraphics[width=2in,height=1in]{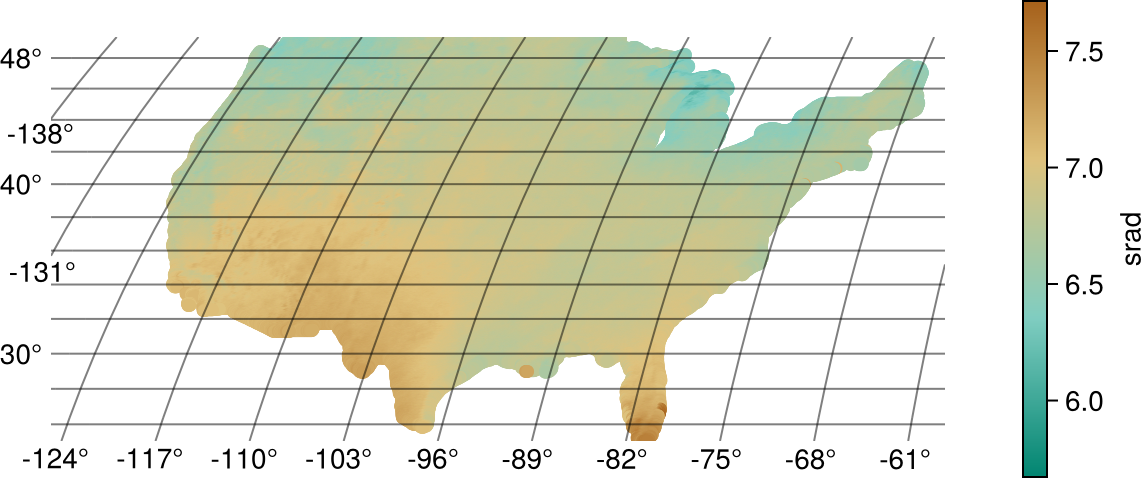}

    \caption{Column 1: gridMET mean. Column 2: TEASE mean.}
    \label{fig:sd_TEASE}
\end{figure}

\begin{figure}[h!]
    \centering
    \includegraphics[width=2in,height=1in]{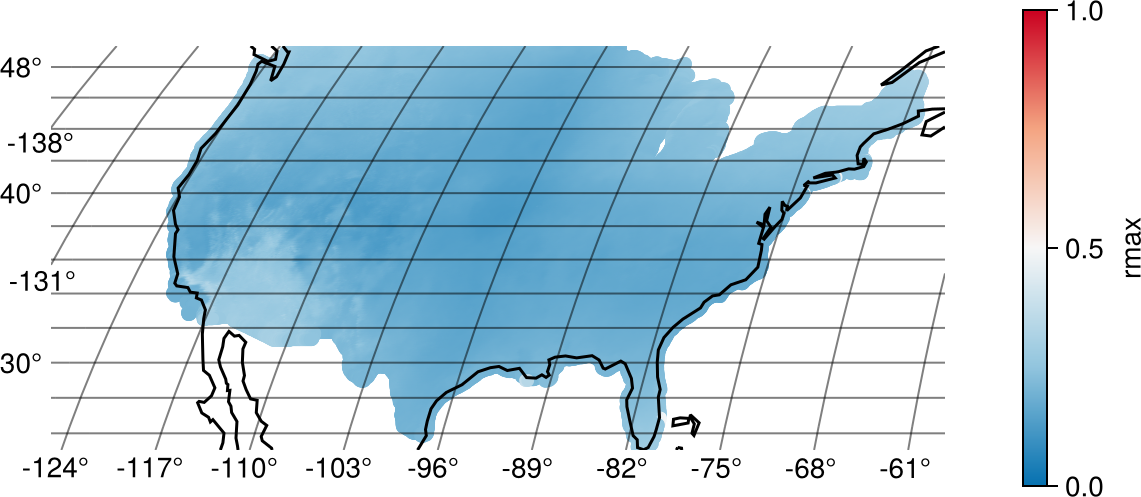}
    \includegraphics[width=2in,height=1in]{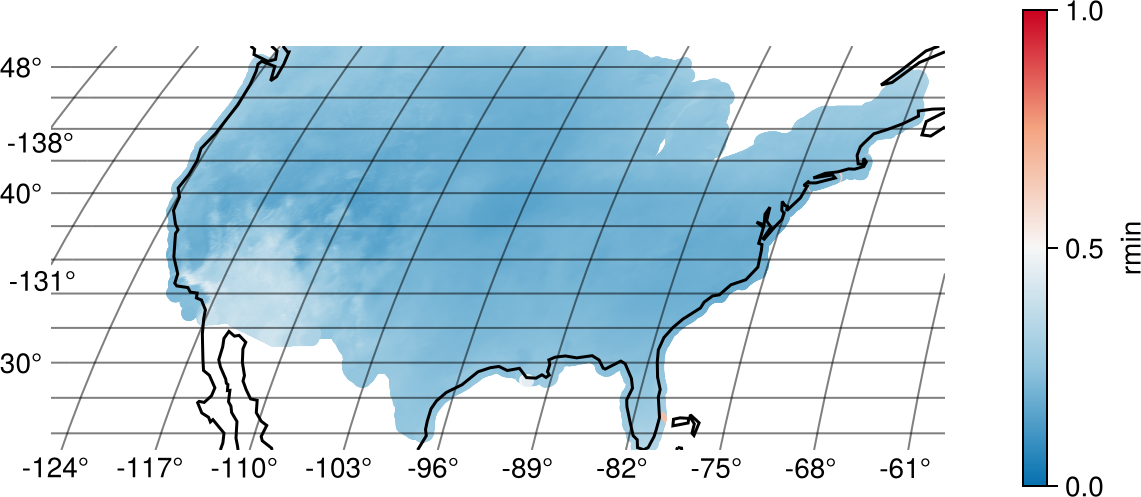}

    \includegraphics[width=2in,height=1in]{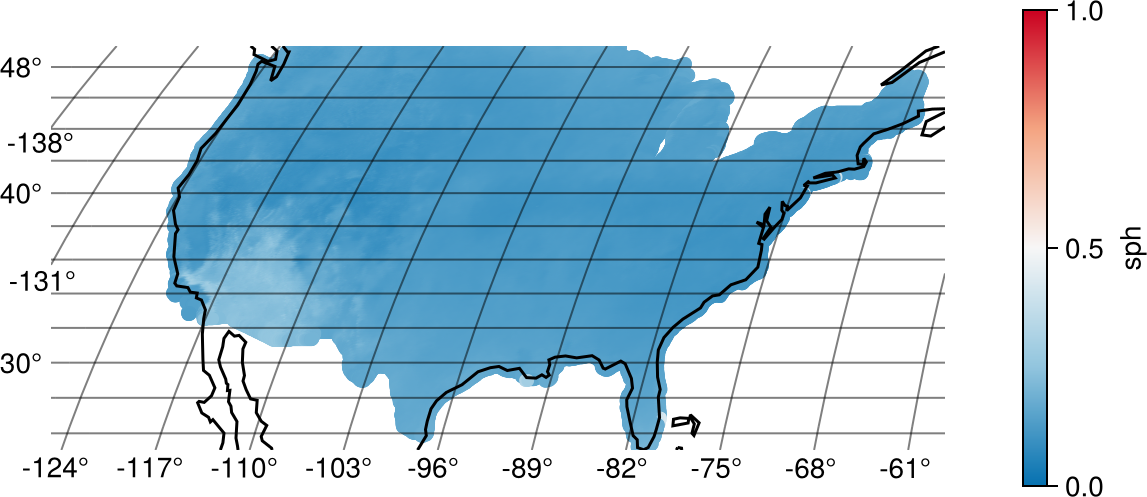}
    \includegraphics[width=2in,height=1in]{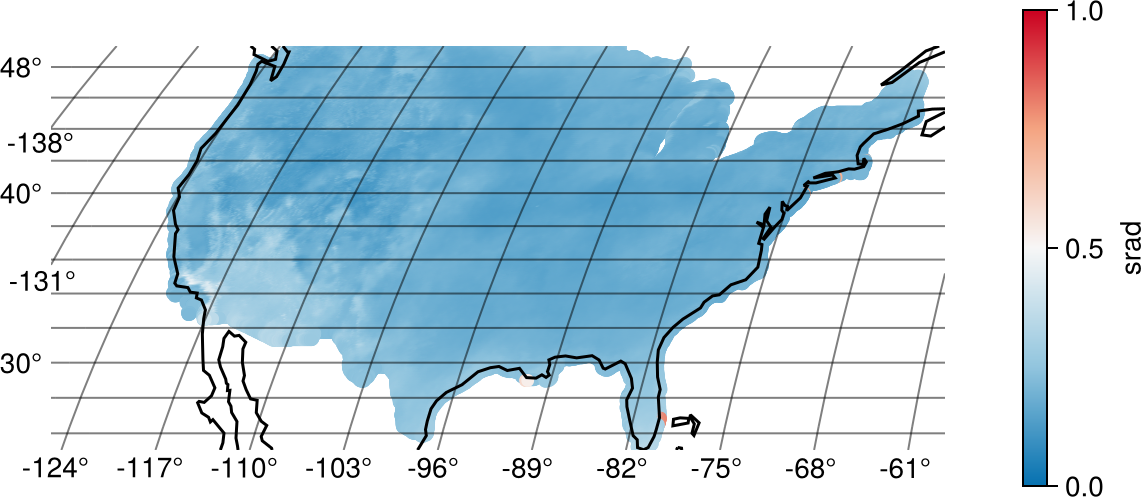}

    \caption{Local TEASE standard deviations.}
    \label{fig:sd_metdata}
\end{figure}

\begin{figure}[h!]
    \centering
    \includegraphics[width=2in,height=1in]{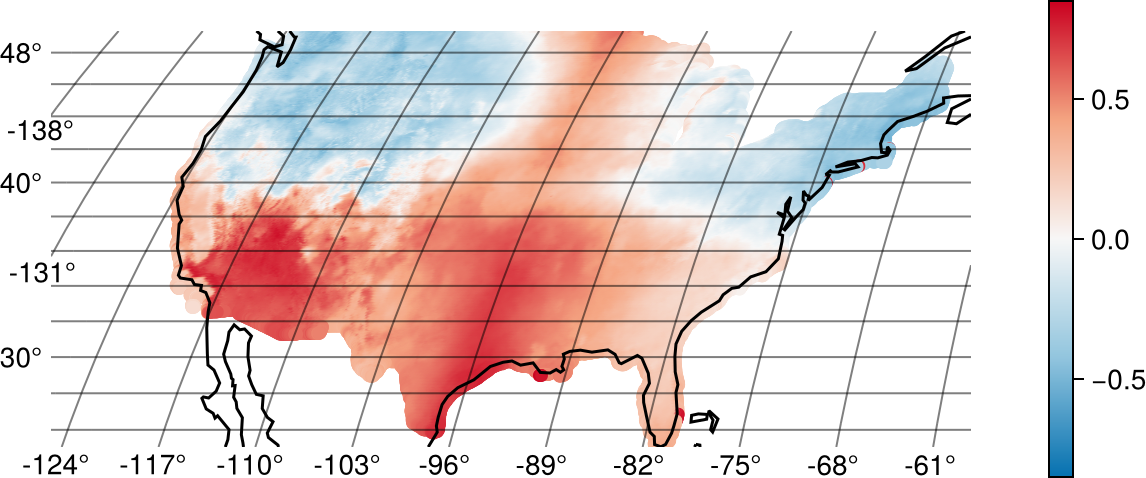}
    \includegraphics[width=2in,height=1in]{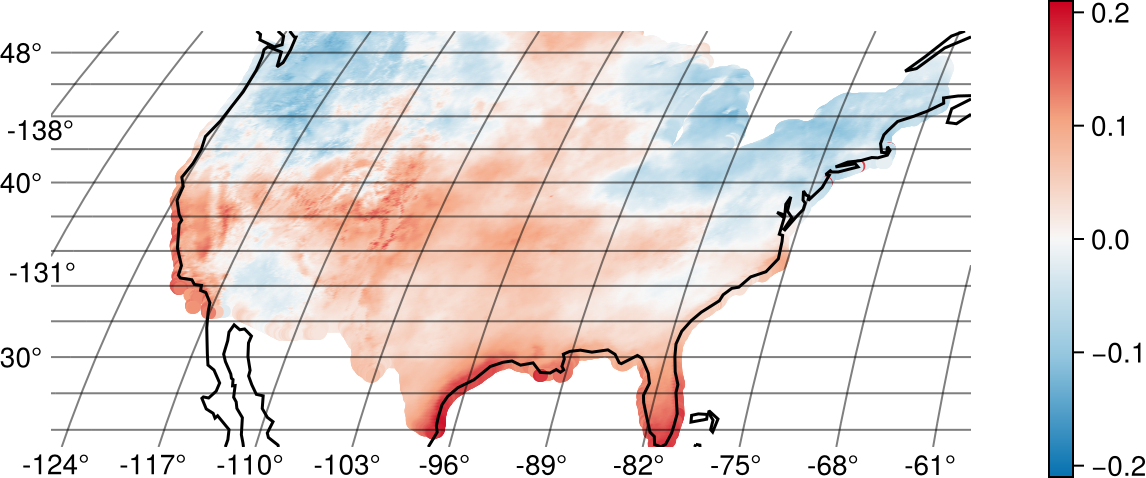}

    \caption{Local TEASE cross-correlations between \texttt{rmax} and \texttt{rmin} (Column 1) and \texttt{rmax} and \texttt{srad} (Column 2).}
    \label{fig:sd_metdata}
\end{figure}

\section{Conclusion}
\label{sec:discussion}

TEASE is a new multi-task GP, flexibly splitting between efficient updates for immaterial and material parameters.
Full-scale BGL \citep{leduc2025modelinglargenonstationaryspatial} inspires basis selection: EOFs for material, and compactly-supported multiresolution basis functions for immaterial.
Critically, envelope orthogonality lets us sidestep semi-orthonormality of immaterial basis and leverage fixed rank co-kriging \citep{caringi2026fixedrankcokrigingmodel}.
To mitigate bias, immaterial covariance is penalized \citep{bien} rather than precision.
Meanwhile, targeted graphical elastic net \citep{kovács2021graphicalelasticnettarget} regularizes material precision matrices.
Enhanced penalty on regression coefficients \citep{kwon2024} leverages envelope mechanics to improve prediction risk and solve identifiability issues.
\cite{widjaja2026} recently studied spatio-temporal LCM partial envelope.
Besides a time extension, natural TEASE targets are nonstationary SAR \citep{wiens2020} and Bayesian \citep{PAIGE2022107503}.

\section{Acknowledgments}

Thanks to Tim Duff for useful conversations about Stiefel manifold optimization.

\section{Disclosure statement}\label{disclosure-statement}

The authors have no conflicts of interest.

\section{Declaration of Generative AI Use}\label{AI-statement}

No generative AI was leveraged for research, coding, or writing. 

\section{Data Availability Statement}\label{data-availability-statement}

Data can be downloaded at \url{https://gdex.ucar.edu/dataset/371_abaker.html} and \url{https://www.climatologylab.org/gridmet.html}.

\begin{center}

{\large\bf SUPPLEMENTARY MATERIAL}

\end{center}

Code is available at \url{github.com/mlkrock/TEASE}.

  \bibliography{envelope}
\bibliographystyle{agsm}

\end{document}